\documentclass[trackchanges, twocolumn]{aastex701}

\begin{document}

\title{Baryon-Accelerated Core Collapse in SIDM Halos and its Imprint on Galactic Disks}


\author[0009-0003-6359-5603]{Zeineb Mezghanni}
\affiliation{Department of Physics and Kavli Institute for Astrophysics and Space Research, MIT, Cambridge, MA 02139, USA}
\email{zeinebm@mit.edu}

\author[0000-0001-5996-4072]{Elliot Y. Davies}
\affiliation{Department of Physics and Kavli Institute for Astrophysics and Space Research, MIT, Cambridge, MA 02139, USA}
\email{eydavies@mit.edu}

\author[0000-0002-7352-6252]{Adriana Dropulic}
\affiliation{DARK, Niels Bohr Institute, University of Copenhagen, 2200 Copenhagen, Denmark} 
\email{adriana.dropulic@nbi.ku.dk}

\author[0000-0001-9250-8597]{Gonzalo Herrera}
\affiliation{Department of Physics and Kavli Institute for Astrophysics and Space Research, MIT, Cambridge, MA 02139, USA}
\affiliation{Harvard University, Department of Physics and Laboratory for Particle Physics and Cosmology, Cambridge, MA 02138, USA}
\email{gonzaloh@mit.edu}

\author[0000-0002-5560-8668]{Abdelaziz Hussein}
\affiliation{Department of Physics and Kavli Institute for Astrophysics and Space Research, MIT, Cambridge, MA 02139, USA}
\email{abdelh@mit.edu}

\author[0000-0003-2806-1414]{Lina Necib}
\affiliation{Department of Physics and Kavli Institute for Astrophysics and Space Research, MIT, Cambridge, MA 02139, USA}
\email{lnecib@mit.edu}

\begin{abstract}
The gravitational coupling between dark matter (DM) halos and the baryonic structures they host is one of the most powerful windows into the particle nature of DM.
Self-interacting dark matter (SIDM) presents a minimal, well-motivated extension to the dark sector with dramatic consequences for the structure of galaxies and their halos. 
However, the impact of baryons on SIDM halo evolution and the resulting galactic structure has been underexplored in Milky Way (MW)-size galaxies. 
In this paper, we demonstrate that the inclusion of a baryonic component in a MW-size galaxy causes accelerated core collapse to begin within the MW's lifetime for a cross section as low as $\sigma/m = 1 \, \rm{cm}^2/\rm{g}$.
We present a suite of $N$-body simulations of cold dark matter and SIDM MW-size galaxies with and without a baryonic component for cross sections $\sigma/m = \left[1.0, 2.5, 5.0\right]$ cm$^2$/g. 
We find numerically, and semi-analytically, that the presence of a stellar disk and bulge shortens the predicted core collapse timescales from the DM only simulations by a factor of $\sim 40$. 
Further, as the core collapse begins within the lifetime of the galaxy, the subsequent density increase strengthens the mid-plane restoring force exerted on stellar disk orbits, leading the disk to flare.  
This work quantifies both directions of the baryon–SIDM coupling: baryons accelerate core collapse in MW-sized halos, and the resulting halo evolution reshapes the disk through thinning and flaring.
Both processes open new observational windows into DM.

\end{abstract}


\section{Introduction} 

Cold dark matter (CDM) predicts approximately universal spherically averaged halo density profiles in dark matter only~(DMO) simulations~\citep{Navarro1996,Navarro1997}. Navarro--Frenk--White (NFW) density profiles have an inner density that scales as $\rho \sim r^{-1}$, producing a central \textit{cusp}, and an outer density that asymptotes to $\rho \sim r^{-3}$. Their velocity dispersion profiles are correspondingly not isothermal: the dispersion rises from the center, peaking within the characteristic scale radius of the NFW, and declines at outer radii \citep{Lokas2001}, making the central regions of CDM halos dynamically colder than their surroundings \citep{Banik_2025}.

Yet a range of observations, such as cored dwarf rotation curves~\citep{Flores_1994, Moore1994, de_Blok_2001, Gentile_2004,simon2005, de_Naray_2008}, the diversity of rotation curves at fixed maximum velocity~\citep{Oman_2015, Santos_Santos_2018, Ren_2019, Garrison-Kimmel_2019, Jiang_2019, Santos_Santos_2020, Sales_2022}, and the apparent absence of dense massive subhalos predicted around the Milky Way (MW)  \citep{Boy_Kol_2011, Boy_Kol_2012, Tollerud_214, Gar_Kim_2014, kirby_2014, Papastergis_2015}, suggest that the inner structure of galactic halos is more complex than predicted by DMO CDM simulations \citep[e.g.,][and references therein]{Bullock_2017}. 

Crucially, however, the inner dark matter distribution in halos containing galaxies does not evolve independently of baryons. Gas cooling, star formation, and stellar feedback can substantially reshape the central gravitational potential. It has been shown in simulations that the accumulation of baryons in the center can contract the dark matter halo \citep{blumenthal, Gnedin:2004cx}. On the other hand, feedback can transform cusps into cores in dwarf galaxies ~\citep{Navarro1996, Mashchenko2008, Governato2010,Governato2012,  Pontzen2012, Brooks2014, Madau_2014, Onorbe2015, Read_2019}. These effects are mass-dependent: feedback-driven core formation becomes less efficient both in ultra-faint galaxies $(M_* / M_{\rm halo} \lesssim 0.01)$ and in galaxies with a high baryonic fraction $(M_* / M_{\rm halo} \gtrsim 0.5)$, where adiabatic contraction dominates \citep{DiCintio2014, Tollet2016, lazar2020}. The structure of the inner halo is therefore directly dependent on the interplay between dark matter and baryons. 

This connection is particularly important when considering self-interacting dark matter (SIDM), which can resolve the aforementioned small-scale structure problems while preserving the successes of CDM on large scales~\citep{Carlson1992, Spergel_2000}. In this paradigm, DM particles elastically scatter, producing thermal conduction from the outer halo towards the inner colder core. This thermalizes the center and transforms the density cusp into an approximately flat \textit{core} $\rho \sim r^0$ in a process known as \textit{core formation} or \textit{core expansion}~\citep{Vogelsberger2012, Rocha2013, Kaplinghat2016, Tulin2018, Ren_2019, Robertson2019}. Because a self-gravitating system has a negative heat capacity, the very conduction that builds a core eventually causes it to collapse. Once the thermalized core is hotter than the surrounding envelope, the net direction of heat flow reverses toward the outer halo, the center contracts and heats, and the halo undergoes a runaway gravothermal collapse toward a new, denser cusp \citep{Balberg_2002, Koda_2011}. SIDM halos thus follow a characteristic cusp--core--cusp evolution, and the late collapse phase, arguably the most distinctive SIDM signature, is reached on a timescale set by the SIDM scattering rate and the halo structure \citep{Outmezguine:2022bhq, Palubski_2024}.

Recent analytic work, semi-analytic work, and DMO $N$-body simulations suggest that, in the absence of baryons, the timescale of core collapse for MW-sized halos exceeds the age of the universe \citep[e.g.][]{Essig_2019, Sameie_2021}. The inclusion of central baryonic components to galactic halos dramatically changes this picture. After eons of accretion and secular inflow, large galaxies like the MW build up a dense inner region of baryons which dominate the central potential \citep[][]{Kormendy2004Secular, Bland_Hawthorn_2016, Portail_2017}. 
The deepened baryonic component contracts and heats the inner halo, thereby increasing the SIDM scattering rate. $N$-body simulations have shown that a sufficiently massive or centrally concentrated baryonic component can substantially shorten the core-expansion phase and drive MW-mass SIDM halos into the core-collapse phase, including for cross sections as low as $\sigma/m=1$ ${\rm cm^2g^{-1}}$~\citep{Elbert_2018, Sameie_2018, Robles_2019}. Semi-analytic calculations similarly find that, for a $M_{\rm vir}=10^{11} M_\odot$ halo with $\sigma/m=1$ ${\rm cm^2g^{-1}}$, the onset of collapse depends strongly on the mass and compactness of the baryonic component~\citep{jiangSemianalyticStudySelfinteracting2023}. Cosmological simulations of MW-mass halos with $\sigma/m=1$ and $10~{\rm cm^2g^{-1}}$ further highlight the importance of baryonic contraction in setting the inner density structure of SIDM halos~\citep{Sameie_2021}. More recently, $N$-body simulations have shown that a sufficiently compact stellar component can cause MW-mass SIDM halos to bypass the usual core-formation phase and enter core collapse even for $\sigma/m=1$ ${\rm cm^2 g^{-1}}$ ~\citep{zengBypassedCoreFormation2026}, an order of magnitude below the upper limit $\sigma/m \leq 10$ ${\rm cm^2 g^{-1}}$ for the MW \citep{Correa_2021}, and substantially below the cross sections typically required for collapse in DMO halos. Despite these advances, work remains to understand the role of baryonic geometry in setting the gravothermal evolution and core-collapse timescale. While the previous studies have primarily varied the mass and compactness of the baryonic component, including the scale length of imposed disks~\citep{Sameie_2018}, this work focuses specifically on an extended disk geometry, deriving its expected impact on the core-collapse timescale and testing this prediction numerically against DMO MW-mass SIDM halos.

Evident by the dramatic role that DM halos play in shaping and evolving large stellar structure \citep[e.g.][]{Beane_2023, dattathriSelfinteractingDarkMatter2026}, the impact that the DM has on baryons requires equal attention as the converse. More precisely, the impact of \textit{gravothermally evolving} SIDM halos on large galactic structure is important yet under-studied ~\cite[see][and references therein for some details]{Adhikari_2025}. As the SIDM halo undergoes gravothermal evolution, it can considerably restructure the baryonic distribution. \cite{zengBypassedCoreFormation2026} show that the stellar component contracts as the central SIDM density grows, with the reduction in stellar half-light radius over time increasing with self-interaction cross section (their Fig.~3c). Crucially, they consider only a spherically symmetric stellar distribution and note that the effects of a non-spherical disk geometry on SIDM gravothermal evolution have not yet been rigorously quantified. We consider such a non-spherical stellar geometry in this work.

The response of the disk's vertical structure to the core-collapsing SIDM halo has not yet been studied or quantified in the literature, which is a direct channel through which the impact of the SIDM halo on the baryonic distribution can be studied and observed. The vertical scale height of a stellar disk is set by the balance between stellar random motions and the local gravitational restoring force, which the central DM impacts. As gravothermal contraction increases the central dark matter density, the disk can respond vertically to the restoring force. Unlike previous studies that approximate the baryonic component as spherical or focus on the instabilities within the disk, in this work we study the vertical structure of a live, flattened stellar disk throughout the host halo's gravothermal evolution. This allows us to quantify how the geometry of a MW-like disk impacts the SIDM core-collapse process as well as identify the imprint that this evolution leaves on the disk's profile, opening up a new observational window into DM physics. Additionally, because disk heating and stability influence star formation and chemical enrichment, these effects may leave observable structural and chemical imprints in the disk~\citep[see e.g.,][]{Leroy_2008, Sch_nrich_2009,minchevChemodynamicalEvolutionMilky2013,mackerethDynamicalHeatingMilky2019,matteucciModellingChemicalEvolution2021,cerquiChemicalEnrichmentHistories2025}. 

In this work, we isolate the co-evolution of the stellar disk and its self interacting halo with controlled $N$-body simulations of a MW--mass galaxy, evolved in isolation for CDM and SIDM halos with $\sigma/m~=~1.0,\,2.5,\,5.0~{\rm cm^2\,g^{-1}}$, each realized with and without a baryonic bulge and disk. First, we demonstrate gravothermal core collapse in MW--size halos, an evolution followed until now almost exclusively at dwarf scales. Second, we show that a baryonic disk enhances the adiabatic contraction of the halo and accelerates its collapse, bringing MW--size halos to begin the core collapse phase within a Hubble time for cross sections as low as $\sigma/m = 1~{\rm cm^2\,g^{-1}}$, whereas their DM-only counterparts do not collapse within the age of the universe. Third, we introduce a simple analytic model for the coupled disk-halo system that characterizes this accelerated collapse and isolates the role of disk geometry from the role of the total enclosed baryonic mass. Fourth, we characterize the observable correlations it predicts: the collapsing halo thins the inner regions of the disk causing it to flare out at larger radii. This effect is more drastic with increasing cross section, tying $\sigma/m$ directly to the radial profiles of galactic disks. This connection therefore provides new cosmological utility for statistical studies of disk-hosting galaxies. This paper is structured as follows: Section~\ref{sec:sims} describes the simulations; Section~\ref{sec:corecollapse} presents the accelerated core collapse and its analytic interpretation; Section~\ref{sec:disk} quantifies the disk thinning and flaring; and Section~\ref{sec:conclusions} concludes.

\section{Simulations}\label{sec:sims}
In this work, we generate a MW-size galaxy using the \texttt{GALIC} code \citep{Yurin_2014}, that we then evolve for 10 Gyr for CDM and three SIDM cross sections using the 2022 version of the public \texttt{GIZMO} package \citep{Gizmo} \footnote{\url{https://bitbucket.org/phopkins/gizmo-public}}.

More specifically, we use the \texttt{GALIC} code to model the DMO galaxy as a \cite{Hernquist1990} DM halo of mass $M_{\rm DM}= 9.90 \times 10^{11} {\rm M}_\odot$. We model the Halo+Baryon galaxy as an exponential stellar disk of mass $M_{\text{disk}}~=~1.20 \times 10^{10} {\rm M}_\odot$ and a  Hernquist bulge of mass $M_{\text{bulge}}~=~2.90 \times 10^{10}  {\rm M}_\odot$ embedded in a \cite{Hernquist1990} DM halo of mass $M_{\rm DM}=  9.90 \times 10^{11} {\rm M}_\odot$. Our simulations do not contain a gas component and do not account for the stellar evolution and the growth in the MW. Spurious collisional heating by finite-mass dark matter particles can artificially increase the velocity dispersion and thickness of simulated stellar disks~\citep{Ludlow_2021}. Our MW host is resolved with $\simeq1.5\times10^7$ dark-matter particles and $6.3\times10^5$ stellar particles, placing the halo well above the $\sim10^6$-particle scale below which such numerical heating becomes substantial for MW-mass systems. We use a particle resolution\footnote{This resolution is chosen close to that of the \emph{Latte} suite~\citep{Wetzel2016} from Feedback In Realistic Environment (FIRE-2) suite~\citep{Hopkins2018} that has similarly used the \texttt{GIZMO} code. In \emph{Latte}, the DM particle mass is $\sim 3.5\times 10^4~M_{\odot}$ while the stars have an initial mass of $\sim 7070~M_{\odot}$.} of $6.56 \times 10^4 {\rm M}_\odot$. \texttt{GALIC} generates a random distribution of all particles in the galaxy and iteratively modifies their positions and velocities to produce a stationary solution to the collisionless Boltzmann equation. 

While our bulge is more massive than the dynamical estimate of the MW bulge, $\sim 1.5-2.0 \times 10^{10} {\rm M}_\odot$ \citep{Bland-Hawthorn_2016,Portail_2017}, only $\sim 36\%$ of our simulated bulge $M_{\text{bulge}}$ ($\sim 1.2 \times 10^{10} {\rm M}_\odot$) is enclosed within $3$~kpc. The remainder of the mass lies at larger radii and is treated as an effective central baryonic potential of the halo. This compensates for the gas mass ($\sim 10^{10} {\rm M}_\odot$), which is commonly omitted in collisionless $N$-body simulations \citep{Bland_Hawthorn_2016} . 

We use the \texttt{GIZMO} package to evolve the generated galaxy with CDM and SIDM. We sweep over isotropic velocity-independent cross sections of $\sigma/m~=~1.0,  2.5 , \text{ and } 5.0 \text{ cm}^2/\text{g}$, labeling the respective simulations as SIDM $1$, SIDM $2.5$ and SIDM $5$ throughout this work. 
Self-interactions are implemented in \texttt{GIZMO} using the Monte-Carlo pairwise-scattering scheme described in \cite{Rocha2013}. Each DM particle is assigned an adaptive kernel radius $h_i$ enclosing a fixed effective number of neighbors. If the kernels of a DM pair overlap ($r_{ij} < h_i + h_j$) over a timestep $\Delta t$, they scatter with a probability $P_{ij}$ isotropically in their center-of-mass frame, conserving energy and momentum.

Gravothermal collapse introduces numerical challenges that
must be addressed. Recent convergence studies of idealized SIDM simulations ~\citep{Palubski_2024, mace2024convergencetestsselfinteractingdark, Fischer_2024} and cosmological simulations \citep{engelhardt2026marvelouslydarkdensityprofile,silverman_2026}, show that default or CDM-motivated parameter choices that set the timestepping rate can be insufficient for core collapse, where inadequate timesteps can stall or reverse the core evolution. To ensure the scattering is resolved in time, we conduct convergence tests to justify our parameter choices, which we discuss in Appendix~\ref{app:convergence}.

\section{Evolution of the halo}\label{sec:corecollapse}

 \begin{figure*}[t]
    \centering
    \includegraphics[width=0.85\textwidth]{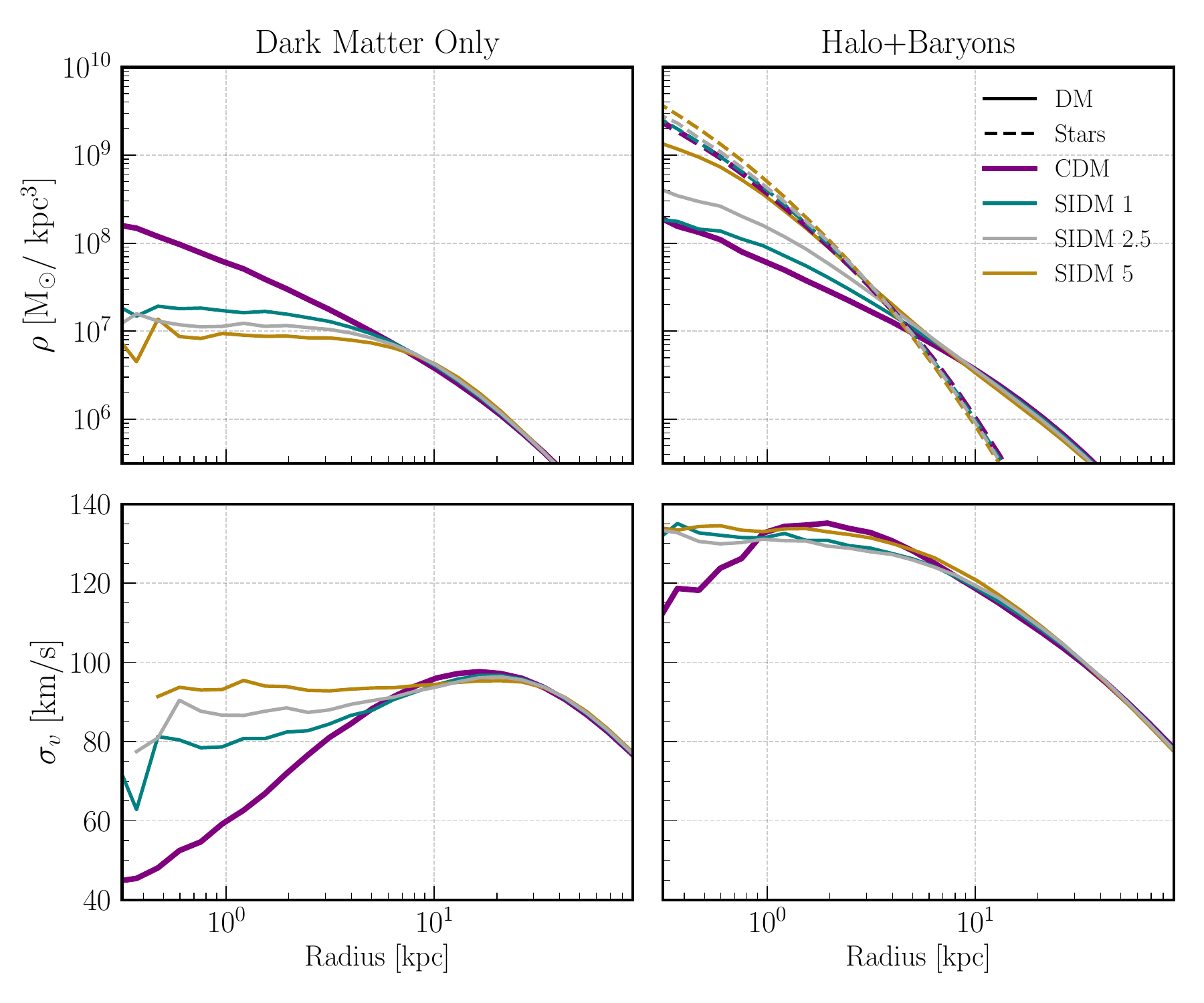}
    \caption{Total density (top row) and velocity dispersion (bottom row) for the DM only (left column) and the DM with baryons (right column) simulations at present time.}
    \label{fig:densitiesanddispersion}
\end{figure*}

In this section, we detail the consequences of the inclusion of baryons for the gravothermal evolution of SIDM halos. First, we explore this impact on the final density and velocity dispersion profiles. We then calculate and compare the core expansion ($t_{\rm core}$) and core collapse ($t_{\rm coll}$) timescales analytically with those found in our simulations. 

\subsection{Impact of baryons on density profiles}\label{sec:coreexp}

\begin{figure*}[]
    \raggedleft
    \includegraphics[width=\textwidth]{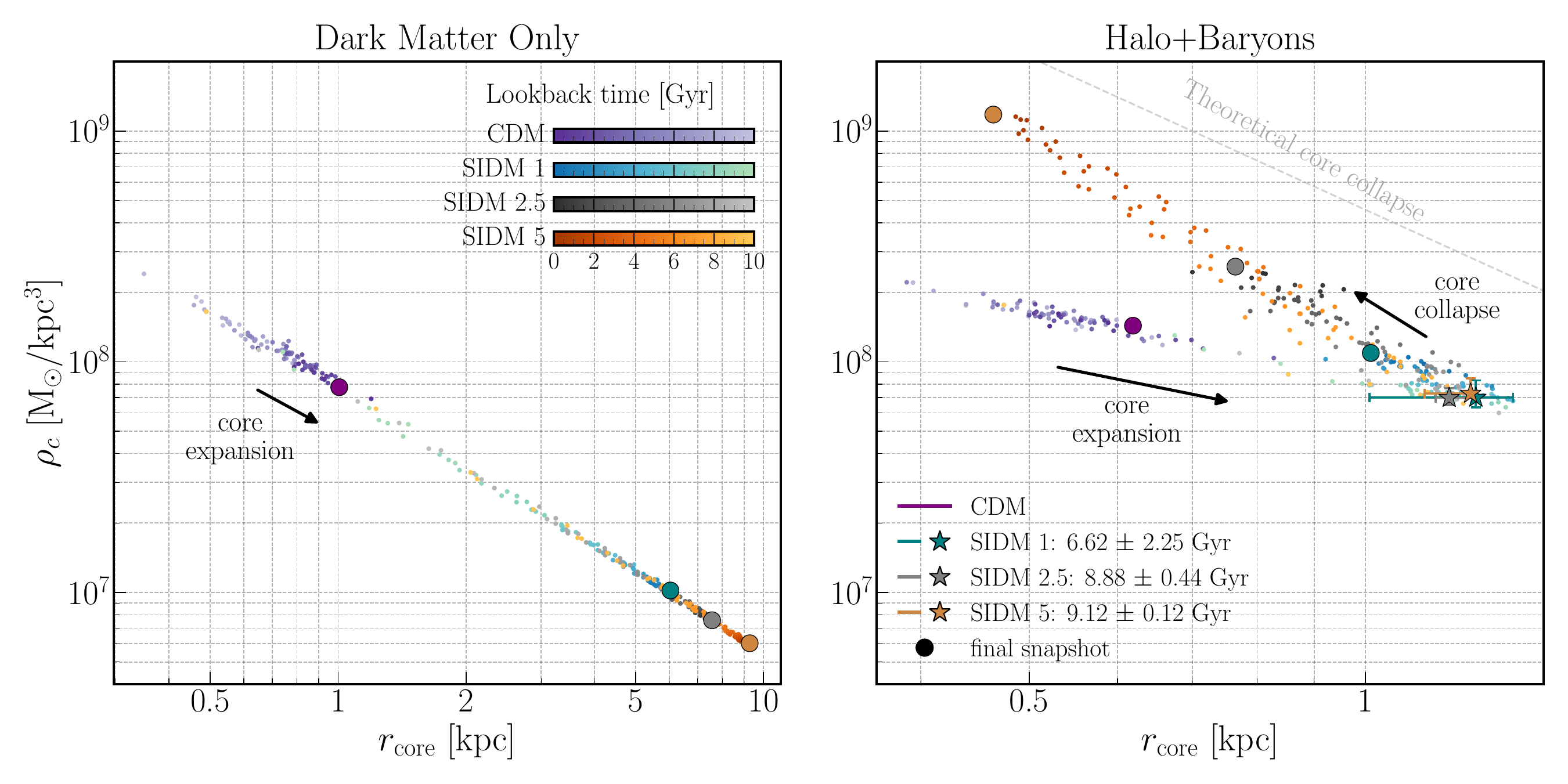}
    \caption {Parametric plot of the core density $\rho_{\rm c}$ vs the core radius $r_c$ as a function of lookback time for the different cross sections. DMO simulations are shown to the left and the Halo+Baryon simulations are to the right. The stars indicate the core density and radius at the end of the core expansion phase and the bigger points depict the final snapshot of the simulation. The error bars on the time of maximal core expansion arise from the fitting procedure described in Sec.~\ref{sec:corecollapsetimesinsim}. We show the theoretical prediction for the core collapse phase with the dashed gray line \citep{Outmezguine:2022bhq}. 
    \label{fig:tc}}
\end{figure*}

In this subsection, we present the density profiles and velocity dispersions of our CDM and SIDM halos in isolation and with a baryonic component. We show that the inclusion of baryons reverses the direction of heat flow in the inner SIDM halo and sets up the accelerated collapse (quantified in Section~\ref{sec:densityincrease}). 

In Figure~\ref{fig:densitiesanddispersion}, we show the present day DM density profiles for CDM and the different SIDM cross sections (top row) and their velocity dispersions (bottom row) of the DMO (left column) and Halo+Baryons (right column) runs. We also show the stellar density profiles by dashed lines. All runs within a column have the same initial conditions, so the differences between the curves are entirely due to the DM self-interaction. 

The left column of Figure~\ref{fig:densitiesanddispersion}, shows the DMO runs. The SIDM DMO halos form progressively larger cores with increasingly isothermal velocity dispersion profiles at higher cross sections, compared to the CDM DMO halo. 
In the top-left panel, the CDM halo (purple) retains its cusp and rises as $\rho \sim r^{-1}$ down to the innermost resolved
radius, reaching $\sim 1.5 \times 10^{8}\,{\rm M}_\odot\,{\rm kpc}^{-3}$ at $r \simeq 0.3$ kpc. 
The SIDM halos instead flatten into constant-density cores, with central densities that decrease monotonically with cross section: at the same radius, SIDM~1 (teal) sits $\sim$ an order of magnitude below CDM and SIDM~5 (yellow) roughly a factor of two below SIDM~1. All four curves overlap beyond $r \sim 10$ kpc, where the scattering rate is too low for self-interactions to have redistributed heat over 10 Gyr. 
The bottom-left panel shows the origin of this behavior. The CDM dispersion profile (purple) climbs steeply from $\sim 45\,{\rm km\,s^{-1}}$ in the center to its peak near $r \sim 15$ kpc, so $\partial \sigma_v^2/\partial r > 0$ throughout the inner halo. With increasing SIDM cross section, the inner velocity dispersion profile becomes progressively flatter where $\partial \sigma_v^2 / \partial r \to 0$ out to $r \sim 10$ kpc. The inward heat flow causes the temperature to increase more in the core. Because a self-gravitating system has a negative heat capacity, the inner halo cools and expands leading to a core formation \citep{Spergel_2000, Vogelsberger2012, Rocha2013, Kaplinghat2016, Tulin2018, Ren_2019, Robertson2019}. Figure~\ref{fig:densitiesanddispersion} therefore highlights how this core expansion is more amplified by the scattering cross section where the flatter the inner dispersion profile in the bottom panel, the more the core has expanded in the top panel. 

In the right column of Figure~\ref{fig:densitiesanddispersion}, we find that the inner densities for the SIDM Halo+Baryon simulations are much higher compared to their DMO counterparts. The density enhancement is correlated with an increase in the velocity dispersions $\sigma_v$ (bottom row of Figure~\ref{fig:densitiesanddispersion}) extending to the edge of the stellar disk at $r\sim10$ kpc. This increase in the inner density and velocity dispersion is consistent with what has been seen with the ``Feedback In Realistic Environment" (FIRE) results in \cite{Sameie_2021}.

The enhancement in the velocity dispersions also appears in CDM with the inclusion of the baryons compared to the DMO. This indeed isolates the origin of the inner heating as gravitational contraction by the baryons rather than SIDM thermalization.

When DM dominates the potential, the virial theorem dictates that the velocity dispersion scales as 
\begin{equation}
    \sigma_{v,0}^2(r) \propto \frac{G M_{\rm DM}(<r)}{r},
\end{equation}
where $M_{\rm DM}(<r)$ is the DM mass enclosed within radius $r$ and $G$ is the gravitational constant. When a significant baryonic potential $M_b$ is additionally included \citep[as has been done in][for example]{Zhong_2023}, the velocity dispersion adjusts to
\begin{equation}
    \sigma_v^2(r) \propto \frac{G \left(M_{\rm DM}(<r) + M_{\rm b} (<r)\right)}{r}.
\end{equation}
This can also be written as 
\begin{equation}\label{eq:vdispersion}
    \sigma_v^2(r) \propto \sigma_{v,0}^2\left(1 + \frac{M_{\rm b}(<r)}{M_{\rm DM}(<r)}\right),
\end{equation}
which is a purely virial statement; it depends only on the total enclosed mass, not on its distribution. Since baryons dominate the inner halo, we know that the ratio $M_{\rm b} / M_{\rm DM} $ exceeds unity within the halo's scale radius. Indeed, the bottom panel of Figure~\ref{fig:densitiesanddispersion}  shows that between $r\sim 1$ and $10$ kpc, we find that $\partial \sigma_v^2/\partial r > 0$ for the DMO simulations and $\partial \sigma_v^2/\partial r \leq 0$ in the Halo+Baryons simulations. In the DMO simulations, the positive inner temperature gradient drives heat inward, causing the core to expand. 
However, with the additional heating from a central stellar component, the initial inner temperature gradient is far reduced compared to the DMO case. Therefore, the system is able to reach isothermality much quicker and the core-expansion phase is accelerated.
Once the inner halo becomes isothermal, with a higher temperature than that of the outer halo, the temperature gradient reverses and heat subsequently flows from the core to the colder outer halo.  This outward heat loss forces the core to contract and grow hotter, steepening the temperature gradient and driving a runaway process in which the central density increases without bound and a new, steeper cusp forms \citep{Balberg_2002, Koda_2011}. 
The top panel of Figure~\ref{fig:densitiesanddispersion} shows this impact directly: the central densities of our simulated SIDM halos show cores in the DMO case and rise well above the CDM values in the presence of baryons.

\subsection{Analytical prediction of timescales}\label{sec:densityincrease}

In this subsection, we present an analytic estimate of the core-collapse timescale in the presence of baryons, first in a spherical approximation and then accounting for the disk geometry.  

\subsubsection{Timescales of gravothermal collapse}

The time to core collapse is set by the rate at which self-interactions redistribute heat. At a given radius $r$, the local relaxation timescale reads,
\begin{equation}\label{eq:t_relax}
   \tau (r) = \frac{1}{a\, \rho(r)\, (\sigma/m)\, \sigma_v(r)},
\end{equation}
where $\rho (r)$ is the DM density at that radius, $\sigma_v$ is the one-dimensional velocity dispersion, and $a\simeq \sqrt{16/\pi}$ accounts for a Maxwellian-averaged scattering rate \citep{Balberg_2002}. The collapse itself proceeds over many relaxation times. Self-similar solutions of the gravothermal fluid equations show that the time to reach the singular state is a fixed multiple of $\tau$
\begin{equation}\label{eq:corecollapse}
    t_{\rm coll} \simeq \xi^{\star} \tau,
\end{equation}
with $\xi^{\star} \sim 300-400$ as calculated numerically \citep{Balberg_2002} and validated through numerical simulations \citep{Balberg_2002, Outmezguine:2022bhq, Palubski_2024}, since many scatterings are needed before the cumulative heat loss becomes comparable to the binding energy of the core. Core collapse is set to begin at 

\begin{equation}\label{coreexpansion}
    t_{\rm core} \approx 0.146 \, t_{\rm coll} ,
\end{equation}
which also defines the end of core expansion \citep{Roberts_2025, kong_2025}. 
With the presence of a baryonic disk and bulge, these timescales are shortened. Semi-analytic models of spherical halos by \cite{Zhong_2023} also present how the exact distribution of the baryonic contribution impacts core collapse, finding that assuming a more compact baryonic potential can dramatically shorten the halo core collapse time-scale. Here we estimate the shortening of the core-collapse phase induced by the baryons, first considering a spherical approximation then when accounting for the disk geometry.

\subsubsection{Accelerated core collapse: spherical geometry}

The presence of a baryonic potential is expected to shorten $\tau$, and hence $t_{\rm coll}$, through two compounding effects. First, a deeper central potential raises the velocity dispersion, as shown by Eq.~\ref{eq:vdispersion}, which makes it apparent that the velocity dispersion is enhanced for a sufficiently large baryonic mass $M_{\rm b}$. Second, the baryonic potential draws the halo inward. Under adiabatic contraction, the angular momentum of a circular orbit,
$L^2=G\,M(<r)\,r$, is conserved, so each shell satisfies $r\,M(<r)={\rm const}$ \citep{blumenthal} (see also \cite{Quinlan:1994ed, Gnedin:2004cx, Hussein_2025, Herrera:2026hal} for quantified deviations of this adiabatic invariant). Solving
\begin{equation}
    r_i M_{\rm DM}(<r_i)=r_f\left[M_{\rm DM}(<r_i)+M_{\rm b}(<r_f)\right]
\end{equation}
shell-by-shell yields a contracted profile with an enhanced central density \citep{Zhong_2023}

\begin{equation}
\rho \simeq \rho_0\left(1+x(r)\right).
\qquad
x(r)\equiv\frac{M_{\rm b}(<r)}{M_{\rm DM}(<r)},
\label{eq:rhoc}
\end{equation}
where $\rho_0$ describes the core density in the DMO case and and $x(r)$ is the
local baryon-to-DM mass ratio inside radius $r$. Since $\sigma_v^2$ is likewise enhanced by $(1+x)$ (Eq.~\ref{eq:vdispersion}), the core-collapse timescale is then shortened as
\begin{equation}
    \frac{t_{\rm coll}}{t_0} = \frac{1 / (\rho(r)\, \sigma_v(r))}{1 / (\rho_0(r)\, \sigma_{v,0}(r))} \simeq (1+x)^{-3/2}
\label{eq:analytic}
\end{equation}
where $t_0$ is the DMO core-collapse time.
The cross section per mass $\sigma/m$ cancels out: the relative core collapse time acceleration ($t_{\rm coll}/t_0$) is a function of $x$ alone, while the absolute time scales as $(\sigma/m)^{-1}$. Because the baryons are far more centrally concentrated than the halo, $x(r)$ rises steeply inward: in our simulations $x \in [0.76, 1.05]$ within 10~kpc and $x \in [2.5, 11]$ within $r_{\rm core}$. This scaling suggests why, in the presence of a central baryonic component, our halos begin to core collapse within a Hubble time even at $\sigma/m=1~{\rm cm^2\,g^{-1}}$ \citep{Elbert_2018, Sameie_2018, Robles_2019}, whereas their DMO counterparts remain in the expansion phase throughout. Figure~\ref{fig:tc} shows the core density $\rho_{\rm c}$ as a function of the core radius $r_{\rm core}$. The scatter plot shows the evolution of $\rho_{\rm c}$ and $r_{\rm core}$ as a function of time for each of the models. The stars mark the end of the core expansion phase (equivalently, the onset of runaway collapse) for each SIDM model (fitting method described in Section~\ref{sec:corecollapsetimesinsim}), confirming that the core expansion is accelerated by the presence of the baryons and the increased interaction cross section of the DM particles. 
We also show the final $\rho_{\rm c}$ with the circles for CDM, SIDM~1, SIDM~2.5, and SIDM~5 colored respectively in purple, teal, gray, and yellow.

\begin{figure}[]
    \centering
    \includegraphics[width=\columnwidth]{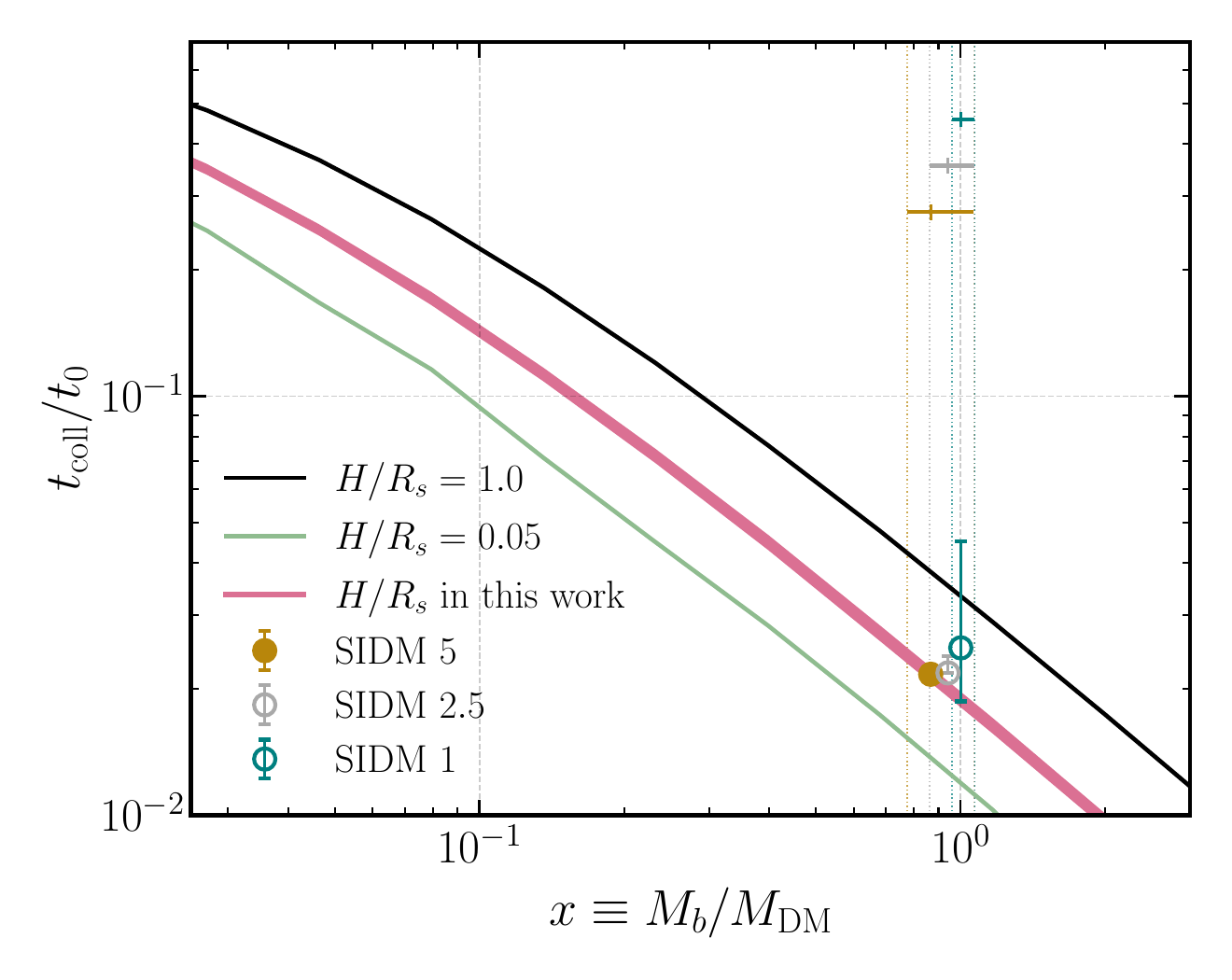}
    \caption{Poisson equation solution for disk geometry. We show a solution for a thick disk ($H/\ R_s=1$) in black, a thin disk ($H/\ R_s=0.05$) in green, and the range for the disks in our simulations ($H/\ R_s=[0.380, 0.442]$) in pink. $t_{\rm{coll}}/t_0$ calculated from the simulations is marked by circles colored in yellow for SIDM 5 with $t_{\rm{coll}} = 8.6^{+0.1}_{-0.0}$ Gyr ($t_0 = 398$ Gyr), gray for SIDM 2.5 with $t_{\rm{coll}} = 17.4^{+1.7}_{-0.0}$ Gyr ($t_0 = 796$ Gyr), and teal for SIDM 1 with $t_{\rm{coll}} = 49.9^{+39.6}_{-12.7}$ Gyr ($t_0 = 1990 \pm 1$ Gyr). The filled circles mark a time that was measured using the simulation data and a hollow circle marks a time that was obtained through self-similarity fits. The horizontal bars show the range of $x$ within 10 kpc for of the simulations. The vertical error bars show the uncertainties on $t_{\rm coll}/t_0$. } 
    \label{fig:poisson}
\end{figure} 

\subsubsection{Accelerated core collapse: disk geometry}
The estimate in Eq.~\ref{eq:analytic} treats the baryons only through their enclosed mass $M_{\rm b}(<r)$, as though they were spherically distributed. Real baryons instead settle into a flattened disk, and the geometry matters in two ways that an enclosed-mass argument cannot capture:
\begin{itemize}
    \item At fixed mass a thin disk is more centrally concentrated than a round distribution, so it contracts the halo more strongly.
    \item A flattened potential exerts an additional vertical pull on the DM that a spherical mass of the same enclosed-mass profile does not.
\end{itemize}  
To include both effects we solve for the coupled potential directly, modeling the halo as an NFW profile \citep{Navarro1996} and the baryons as an exponential disk \citep{freeman} of scale radius $R_s$ and scale height $H$. The disk aspect ratio $H/R_s$ is the single parameter controlling the baryon geometry. 
The relationship between the total gravitational potential and the densities is given by Poisson's equation:
\begin{equation}\label{eq:poisson}
    \nabla^2\Phi
    = 4\pi G\left(\rho_{\rm DM}+\rho_{\rm b}\right)
\end{equation}
where $\rho=\rho_{\rm DM}+\rho_{\rm b}$ is the local density. In cylindrical coordinates, Poisson's equation is written as
\begin{equation}\label{eq:poisson_cyl}
        \nabla^2\Phi = \frac{1}{R}\frac{\partial}{\partial R}
    \!\left(R\frac{\partial\Phi}{\partial R}\right)
    +\frac{\partial^2\Phi}{\partial z^2}. 
\end{equation}
The radial (first) term of Eq.~\ref{eq:poisson_cyl} can be re-expressed in terms of the rotation curve $dV_c/dR$,
\begin{equation}
    \frac{1}{R}\frac{\partial}{\partial R}\left(R\,\frac{\partial \Phi}{\partial R}\right)=\frac{1}{R}\,\frac{d V_c^2}{dR},
\end{equation}
which is subdominant near the midplane of a thin disk\footnote{We have checked the relative size of the two terms in Eq.~\ref{eq:poisson_cyl} by solving Poisson's equation for the exponential disk directly. At the mid-plane, the ratio of the radial term (the first term on the right hand side of Eq.~\ref{eq:poisson_cyl}) to the vertical term ($\partial^2 \Phi /\partial z^2$) scales as ${\sim}H/R$, and it is
$\lesssim10\%$ for $R\gtrsim R_s$ in a thin disk ($H/R_s=0.1$), but reaches
$0.4$--$0.6$ at $R\sim0.3$--$0.5\,R_s$ for the aspect ratios of our disks
($H/R_s\simeq0.4$). Keeping only the vertical term propagates to a $\lesssim15\%$ error in
$t_{\rm coll}/t_0$.} 

It vanishes identically for a flat rotation curve, and more generally it is set by the mean enclosed density, $\bar\rho(R)\equiv 3M(<R)/(4\pi R^3)$ within a sphere of radius $R$. A disk tips the balance, since its mass sits within $|z|\lesssim H\ll R$, so the vertical gradient of the potential is set by the local mid-plane density $\rho_0$. For a disk of scale height $H\ll R$ the mass is concentrated near the plane, so $\rho_0/\bar\rho\sim R/H\gg1$ and the vertical curvature dominates, leaving only the vertical component of the Poisson's equation: 
\begin{equation}
\frac{\partial^2 \Phi}{\partial z^2} \simeq 4\pi G(\rho_{\rm DM}+\rho_{\rm b}).
\end{equation}
In this description, the stars behave as a simple harmonic oscillator with frequency $\nu$ \citep{BT2008} set by the local density,
\begin{equation}\label{eq:nuz}
\nu^2 \equiv \frac{\partial^2 \Phi}{\partial z^2}\bigg|_{z=0}
\simeq 4\pi G\left[\rho_{\rm DM}(R,0)+\rho_{\rm b}(R,0)\right].
\end{equation}
The same mass distributed spherically gives only $\nu^2~=~GM(<R)/R^3$, fixed by the enclosed mass. The ratio of the two grows as $R/H$ for a disk of scale height $H$. If we assume that the adiabatic vertical action $J_z\simeq\frac12\nu z_{\rm max}^2$ is conserved\footnote{Over the 10~Gyr of our simulations, we find that the actions $J_r$, $J_\phi$, and $J_z$ (calculated using \texttt{agama} \citep{vasiliev2019agama}) change by $1.4\%$, $0.9\%$, and $0.3\%$ respectively, which we treat at effectively constant.}\label{footnote:actions}, this excess compresses the halo vertically. However, since the vertical and radial motions are decoupled, the
column density $\Sigma(R)=\int\rho\,dz$ remains constant and the central density raises by a factor of $(\nu^2)^{1/4}$  
 to account for the decrease in $z_{\rm{max}}$. More about the compression and flaring of the stellar disk is described in Section~\ref{sec:disk}. 

We solve Eq.~\ref{eq:poisson} numerically on a uniform cylindrical $(R,z)$ grid of $170\times170$ cells spanning $0\le R,z\le 12\,r_s$, where $r_s$ is the scale radius of the halo. The radial term of Eq.~\ref{eq:poisson_cyl} is discretized using the half-cell radii $R_{i\pm1/2}$, which remains regular at the symmetry axis, and the resulting sparse linear system is solved by direct LU factorization. We impose three boundary conditions. Reflection symmetry about
the mid-plane lets us solve for $z\ge0$ only, with $\partial\Phi/\partial z=0$ at
$z=0$. On the symmetry axis we impose regularity, $\partial\Phi/\partial R=0$, where the radial operator reduces to $2\,\partial^{2}\Phi/\partial R^{2}$. On the outer edges we impose a Dirichlet condition taken from the multipole expansion of the potential, truncated at the monopole, $\Phi=-GM/r$ with $M=M_{\rm vir}+M_b$ \citep{BT2008}.
Far from the center the monopole falls as $1/r$ and each successive multipole carries an additional power of $1/r$; because the density is axisymmetric and symmetric about the mid-plane, all odd moments vanish and the leading correction is the disk quadrupole, $q_2=M_b(2H^{2}-3R_s^{2})$, smaller than the monopole at
the boundary by $\sim 8\times10^{-4}$ for $M_b/M_{\rm DM}=0.2$ and by at most ${\sim}4\times10^{-3}$ across the range of $x$ in Figure~\ref{fig:poisson}.
Because the halo density is formally divergent at the origin, it is evaluated at a softened radius $\sqrt{R^{2}+z^{2}+\epsilon^{2}}$ with
$\epsilon=1.5\,\Delta R\simeq0.11\,r_s$.

We then contract the halo adiabatically in response to the disk in two steps. First, a spherical adiabatic contraction conserving $r\,M(<r)$ for the spherically-averaged mass \citep{blumenthal}, solving $r_{f}[M_{\rm DM}(r_{i})+M_{b}(r_{f})]=r_{i}M_{\rm DM}(r_{i})$, and mapping the density with $\rho_{f}=\rho_{i}(r_{i}/r_{f})^{2}\,dr_{i}/dr_{f}$.  
Second, the non-spherical vertical compression through a factor $[\nu^2/(GM(<R)/R^3)]^{1/4}$ measured from the solution. 
We read off the central density $\rho_{\rm c}$ and the Jeans dispersion $v$ to obtain the collapse time through Eq.~\ref{eq:analytic}. Figure~\ref{fig:poisson} shows the resulting ratio $t_{\rm coll}/t_0$ as a function of $x=M_{\rm b}/M_{\rm DM}$ for a range of disk thicknesses, where $H/R_s=1$ is a thick disk (black), $H/R_s=0.05$ is a thin disk (green) and  $H/R_s = [0.380, 0.442]$ is the ratio in our 4 simulations over time (pink).  $t_{\rm coll}/t_0$ approaches unity at small $x$, where the sub-dominant baryons neither contract nor heat the halo; falls by an order of magnitude by $x\sim1$, so that halos which never collapse in the DMO case now collapse within a Hubble time once the disk is present; and shortens further, at fixed $x$, for thinner disks. The leading behavior follows the analytic $(1+x)^{-3/2}$ scaling, with the disk geometry adding a secondary, order-unity enhancement that saturates once the disk dominates the central density. This is consistent with the shortened collapse times described in previous SIDM+baryons simulations \citep{Sameie_2018, Robles_2019, Zhong_2023}. Building on these results, we find that accelerated core collapse depends not only on how much baryonic mass a halo hosts but also the morphology of these baryons, where  galaxies with thinner disks core collapse faster.

\subsection{Core collapse timescales in simulations}\label{sec:corecollapsetimesinsim}

In this subsection we describe the fits and approximations we use to calculate the core expansion and core collapse timescales in the Halo+Baryons simulations and how they compare to the DMO predictions. 

\begin{table*}[]
    \centering
    \renewcommand{\arraystretch}{1.5}
    \begin{tabular}{lcccc}
        \hline
        Simulation & $t_{\rm core,0}$ & $t_{\rm core}$ & $t_0$ & $t_{\rm coll}$ \\
        \hline
        SIDM 1   & $290.54 \pm 0.1$ & $3.38 \pm 2.3$ & $1989.98 \pm 1.0$ & $49.9^{+39.6}_{-12.7}$ \\
        SIDM 2.5 & $116.22 \pm 0.1$  & $1.12 \pm 0.4$ & $795.99 \pm 0.0$  & $17.4^{+1.7}_{-0.0}$ \\
        SIDM 5   & $58.11 \pm 0.0$  & $0.88 \pm 0.1$ & $398.00 \pm 0.0$  & $8.6^{+0.1}_{-0.0}$ \\
        \hline
    \end{tabular}
    \caption{We calculate the theoretical predictions for DMO runs for end of core expansion timescale $t_{\rm core,0}$ and core collapse $t_0$ using Eq.~\ref{coreexpansion} and~\ref{eq:corecollapse}. We identify the end of core expansion $t_{\rm core}$ and infer the end of core collapse $t_{coll}$ for the Halo+Baryons simulations by smoothing and fitting the core densities $\rho_{\rm c}$. The times reported are in Gyr.}
    \label{tab:times}
\end{table*}

In our simulations, the DMO halos continue in the core expansion phase for the entire time span of the simulation (10 Gyr) and never collapse, as we show in Figure~\ref{fig:tc}. The predicted core collapse ($t_0$) and core expansion ($t_{\rm core,0}$) timescales for our DMO halos are calculated using Eq.~\ref{eq:corecollapse} and ~\ref{coreexpansion} with a choice of $r=10$ kpc and presented in Table~\ref{tab:times}.

In the presence of the disk, core expansion ends well within the time of the simulation. To quantify the end of core expansion that we observe in simulations, we first smooth the noise in $\rho_{\rm c}$ using a moving average on $\log(\rho_{\rm c})$ with a window size of 1.25 Gyr (equivalent to 10 snapshots). We then identify $t_{\rm core}$ as the time at which the smoothed curve reaches its minimum ($\rho_{\rm core}$) \footnote{The uncertainty on the found value stems from the presence of a region where the smoothed curve stays within $\epsilon$ of the minimum set by the median absolute deviation of the log-residuals.}. We list the resulting times in Table~\ref{tab:times}. 

We define the core collapse time $t_{\rm coll}$ as the epoch at which the $\rho_{\rm c}$ first reaches $10\,\rho_{\rm core}$ \citep{silverman_2026}. Since only SIDM~5 crosses this threshold within the simulated 10~Gyr, we exploit the self-similarity of gravothermal core collapse described in \cite{Outmezguine:2022bhq} to extrapolate $t_{\rm coll}$ for SIDM~2.5 and SIDM~1. 
After rescaling time to 
\begin{equation}
u=\frac{t -t_{\rm core}}{t_{c}-t_{\rm core}},
\end{equation}
each run begins at $t_{\rm core}\equiv u=0$ and diverges at the finite time $t_{c}\equiv u=1$. We find that the central density of every halo follows a single universal curve, $\rho_{\rm c}(t)~=~\rho_{\rm core}\, S(u)$, with the SIDM cross section only affecting the rate of the evolution through the timescale $t_{\rm c} - t_{\rm core}$.

We parametrize the universal curve as\footnote{Other fits including polynomial of degrees 2–4 and log-parabola were tested. However, they imposed a symmetric turning point that the data did not have and their inferred $t_{core}$ was sensitive to the fitted range and polynomial degree.}
\begin{equation}
  S(u) = \frac{a u^2 + b u +1}{(1 - u)^{\beta}} ,
\end{equation}
where $\rho_{\rm c} \propto (t_{\rm c} - t)^{-\beta}$, with $\beta$ of order unity \citep{Lynden-Bell_1980, Balberg_2002, Outmezguine:2022bhq}, encoding the physical runaway divergence, and the polynomial prefactor $a u^2 + b u +1$ permits a flat minimum at the end of core expansion. 
We calibrate the shape parameters $(a, b, \beta)$ on SIDM~5. For SIDM~1 and SIDM~2.5 we determine the anchor $(t_{\rm core}, \rho_{\rm core})$ as described in Section~\ref{sec:coreexp} (so that $S(0)=1$ by construction), hold the shape parameters $(a, b, \beta)$ fixed at the values calibrated on SIDM~5, and fit the divergence time $t_{\rm c}$ as the single free parameter. The threshold $\rho_{\rm c} = 10\,\rho_{\rm core}$ is then reached at the universal point $u^{\ast}$ defined by $S(u^{\ast})=10$, giving
\begin{equation}
  t_{\rm coll} = t_{\rm core} + u^{\ast}\,(t_{\rm c} - t_{\rm core}).
\end{equation}
We report the ratios $t_{\rm coll}/t_0$ in Figure~\ref{fig:poisson} and in Table~\ref{tab:times}. We also mark the ratio $x$ within $R=10$ kpc in our simulations by the horizontal bars in Figure~\ref{fig:poisson}.   Uncertainties on $t_{\rm coll}$ are obtained from repeating the fit at the edges of the $t_{\rm core}$ window calculated in Table~\ref{tab:times}. Calibration shifts the collapse time prediction by only 1.2 Gyr ($0.15\sigma$) for SIDM~1, confirming the result is not driven by the calibration choice.
The large uncertainty in $t_{\rm coll}/t_0$ for SIDM~1 is due to the slow increase of $\rho_{\rm c}$ past $\rho_{\rm core}$ at 10~Gyr, which causes the high uncertainty on $t_{\rm{core}}$. 

\section{Evolution of the disk}\label{sec:disk}

In the previous section, we discussed the density profile evolution of large ($M\gtrsim 10^{11} {\rm M}_{\odot}$) dark halos under gravothermal core-collapse. We now tackle how a stellar population embedded in such a halo (i.e. a galactic stellar disk) responds to this evolution. Specifically, we investigate how the vertical thickness of the disk responds to the gravothermal evolution of the halo. We first analytically identify how the disk may change under adiabatic density increase of the halo core. In brief, core-collapse increases the midplane restoring force and causes the disk to contract at low radii, and hence flare out at larger radii. We compare this toy model with how the disk evolves in our $N$-body simulations and further quantify the contraction and flaring. As in the previous section, we connect this to the self scattering cross-section of DM.

\subsection{Adiabatic compression}\label{sec:adiabatic_compression}

To explore the coupling between the disk's thickness and the gravothermal evolution of the halo, we need a suitable tracer of disk's vertical extent. Suitable tracers include: a) the maximum vertical height of an orbit, $z_{\rm max}$, and b) the scale height of the disk, $H$. While the latter is a more clear observable, it is a collective property of the whole disk that cannot be straightforwardly linked to the non-uniform evolution of the halo. Therefore, we first explain how $z_{\rm max}$ of a star (on a near-circular orbit) varies with the density of the halo $\rho_{\rm DM}$. In the following subsection (Sec.~\ref{sec:thin_and_flare}), we map out the evolution of the scale height of the entire disk.

In this section, we assume the contribution of the disk to the potential is negligible, so the stars orbit is determined only by the dark halo potential $\rho \simeq \rho_{\rm DM}$. This dark halo is taken to have a flat rotation curve, and to undergo adiabatic evolution (i.e. the potential changes on a timescale comparable to or longer than the star's orbital oscillation period). The assumption of a flat rotation curve is justified within the central core, and the adiabticity is verified by the simulations (see footnote~6). If the disk contributes substantially to the potential, then the equation above is modified to include a disk contribution to $\rho$ (c.f. Eq~\ref{eq:poisson}) which weakens the connection between the natural oscillation frequency (see next paragraph), $\nu$, and $z_{\rm max}$.

For simplicity, we have assumed the disk stars to be on circular orbits. Specifically, we are assuming that the star remains on the same circular orbit, such that its vertical motion can be separated from its radial motion.

Stars on circular orbits just above the Galactic plane---whose orbital vertical extent is sufficiently small that the curvature of the potential does not vary appreciably over the orbit---undergo a restoring force $F_z$, analogous to Hooke's law $(F_z~=~-kz)$, with natural oscillation frequency $\nu$,
\begin{equation}\label{eq:restoring_force}
    F_z = -\frac{\partial\Phi}{\partial z} \approx - \nu^2 z.
\end{equation}
The approximate equality comes from Taylor expanding the potential near $z=0$:
\begin{equation}
    \Phi(R,z) = \Phi(R,0) + \frac{1}{2}\nu(R)^2z^2 + \mathcal{O}(z^4).
\end{equation}
In the Hooke's law analogy, $\nu^2$ plays the role of spring stiffness per unit mass. In Eq.~\ref{eq:restoring_force}, $\Phi$ is the Galactic potential, $z$ is the distance above the plane, and the oscillation frequency is given by,
\begin{equation}
    \nu(R)^2 = \left. \frac{\partial^2 \Phi}{\partial z^2} \right\vert_{z=0}.
\end{equation}
By considering the equations of motion, it is clear that $z$ behaves like a simple harmonic oscillator with amplitude $z_{\rm max}$ if $\nu$ is not time dependent. However, if the restoring force is actually time-dependent, with frequency $\nu(t)$, the system mathematically resembles what may be called a \textit{parametrically driven oscillator} or an \textit{adiabatic oscillator}. In this regime, the stars follow the slow change in their environment since they complete many vertical oscillations as the potential deepens at the mid-plane: $|\dot{\nu}|/\nu^2 \ll 1$. 

The adiabaticity of potential evolution helps us to find a relationship between the vertical frequency $\nu$ and the maximum vertical extent of the stars orbit $z_{\rm max}$. To find an expression for $z_{\rm max}$, consider the vertical energy per unit mass $E_z$ of the star near the mid-plane (with $z$-component velocity $v_z$),
\begin{equation}
    E_z = \frac{1}{2}v_z^2 + \frac{1}{2}\nu^2z^2.
\end{equation}
This vertical energy component relates to the vertical action by $J_z = E_z / \nu$ for a harmonic oscillator, which is constant under adiabatic evolution of the potential. The maximum height $z_{\rm max}$ of the star can be found in terms of energy and oscillation frequency by considering the turning points of its orbit, where $v_z = 0$. Subsequently, this can be expressed in terms of the constant action:
\begin{equation}\label{eq:z_max_definition}
    z_{\rm max }(t) = \sqrt{\frac{2E_z(t)}{\nu(t)^2}} = \sqrt{\frac{2J_z}{\nu(t)}},
\end{equation}
where we have made the time-dependence explicit for clarity. 

To complete the picture, we relate the potential (and therefore the oscillation frequency) to the density using Poisson's equation in cylindrical coordinates (Eq.~\ref{eq:poisson_cyl}). For a flat rotation curve, the radial gradient of the potential vanishes, and we are able to relate the mid-plane density evaluated at $z=0$ directly to the oscillation frequency:
\begin{equation}\label{eq:nu_poission}
    \nu(R)^2 = 4\pi G \rho(R,0).
\end{equation}
We can then combine Eq.~\ref{eq:z_max_definition} and Eq.~\ref{eq:nu_poission} to show that,
\begin{equation}\label{eq:zmax_versus_rho}
    z_{\rm max}(R) \propto \rho(R,0)^{-1/4}.
\end{equation}

Altogether, as the local mid-plane density increases (via core-collapse) the vertical trajectory of the each star is confined into a more compact region. For an ensemble of stars who respond to an adiabatic evolution of the local density, this causes a reduction in the overall scale height $H$ of the disk. Since a greater value of $\sigma/m$ causes a greater increase in central density during the core-collapse phase(as discussed in~\ref{sec:densityincrease}), the scale height measured at similar times during core collapse may be smaller as $\sigma/m$ is increased. 

The question that now arises is the following: how do we explicitly connect the mid-plane density $\rho$ to the cross section? The most meaningful density in the context of SIDM is the density at the core $\rho_{\rm c}$. Assuming the change in the mid-plane density scales as $\rho(R,0) \sim \rho_{\rm c}$, and the scale height traces the $z_{\rm max}$ at the core, we may expect
\begin{equation}\label{eq:H_rhoc}
    H = H_{0} \left(\frac{\rho_{\rm c}}{\rho_{c,0}}\right)^{-1/4},
\end{equation}
where $H_0$ is the scale height of the disk at a time when $\rho_{\rm c} = \rho_{c,0}$. Yet scale height does not capture the more interesting radial picture implicit in Eq.~\ref{eq:zmax_versus_rho}. Evidently, across the radius of the galaxy, both the density $\rho$ and its evolution under core collapse are not uniform. Therefore we must examine how $z_{\rm max}$ evolves as a function of radius, not just at the core. A more dramatic density increase at the core than the outer radii (as expected in gravothermal evolution) would suggest disk flaring as an observable of late-time SIDM halos. We inspect this in the simulations in the following sub-section (Sec.~\ref{sec:thin_and_flare}).

Deviations in our simulations from the idealized behavior are expected for a number of reasons:
\begin{itemize}
    \item Stars are assumed to have their guiding radius $R_{\rm g}$ unchanged by the evolving potential. However, the increasing density may actually drive some stars to different guiding radii, impacting their orbital contraction in the vertical direction. This change in radius would consequently impact the vertical potential they feel.
    \item Outside the adiabatic regime, $J_z$ is no longer conserved.  Whether $J_z$ increases or decreases depends on the phase of the orbit, and the exact time evolution of the potential. We would anticipate the actions for orbits who spend most of their time very near the mid-plane to decrease. However, since $J_z$ isn't expected to decrease for all orbits, the collective behavior of all disk orbits may be to contract less, or to oscillate.
    \item While the flat rotation curve is valid within the core radius, this is not expected to hold at all radii. Therefore, in the simulations, we examine the evolution of $z_{\rm max}$ in bins of guiding radius.
    \item As previously mentioned, scale height $H$ may poorly capture the true thickness of the disk due to its reliance on a functional form. As a robustness test, we consider different functional forms in the next section (Sec.~\ref{sec:thin_and_flare}).
    \item The self-gravity of the disk, assumed to be negligible in this description, will provide an additional contribution to $\rho$, weakening the impact that the evolution of $\rho_{\rm DM}$ has on $z_{\rm max}$.
\end{itemize}

Despite the limitations, this toy model is a useful framework for the orientating ourselves through simulation results presented in the next sub-sections.
 
\subsection{Disk thinning and flaring in simulations}\label{sec:thin_and_flare}

In this section, we examine how the inclusion of self-interactions in our simulations has impacted the vertical structure of the disk, with comparison to the expected relationship from an adiabatically evolving flat rotation curve (Eq.~\ref{eq:zmax_versus_rho}). We find that with an increased cross section $\sigma/m$, the scale height $H$ of the disk decreases, and the median $z_{\rm max}$ of stellar orbits decreases across the disk. Additionally, we find that stars with low guiding radii exhibit a more dramatic decrease in $z_{\rm max}$ than those with larger guiding radii, leading to a flared disk. 

\begin{figure}[t]
    \centering
    \includegraphics[width=\columnwidth]{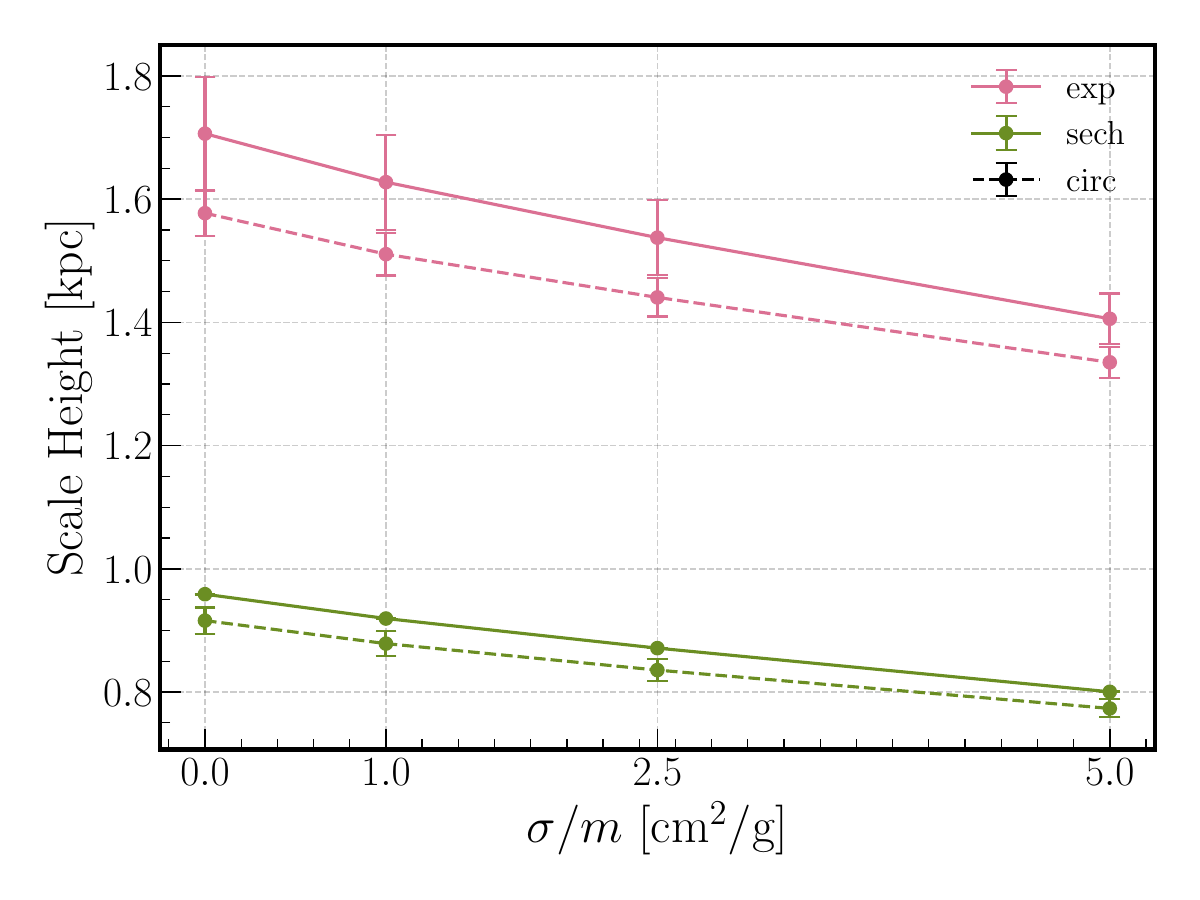}
    \caption{Scale height of the disk as a function of self-interaction cross section at present day. We vary the aperture to be $a = [3, 5.0, 7.5, 10.0, 12.5, 15.0]$ kpc for the exponential (pink) and sech (blue) fits of the disk. We also vary the stars included in the fits to have circularities $\epsilon = [0.2, 0.3, 0.4, 0.5, 0.6, 0.7]$ and fit them to an exponential profile (green). We show the mean and standard deviation for the fitted scale heights as vertical error bars.}
    \label{fig:scaleheights}
\end{figure}

We fit the disk stars of our simulations, initialized by the parameters detailed in Section~\ref{sec:sims}, in two forms:
\begin{equation} 
\rho_*(z) \propto \exp{\left(-\frac{|z|}{H}\right)}
\end{equation}
and
\begin{equation} 
\rho_*(z) \propto {\rm sech}^2{\left(\frac{|z|}{2H}\right)}
\end{equation}
where $z$ is the $z$-component of the stars' position and  $H$ is the scale height of the disk. 
We select the stars to include in the fits using two different techniques: vertical cuts, $|z|<a$, with apertures $a~=~[2.5, 5.0, 7.5, 10.0, 12.5, 15.0, 20.0]$ kpc as well as circularity, $\epsilon$, defined as the ratio of the angular momentum in the z direction of a given star to that of a circular orbit with the same energy \citep{Abadi_2003}. We use the cuts $\epsilon = [0.2, 0.3, 0.4, 0.5, 0.6, 0.7]$ which match the $\epsilon$ range for the thick disk in \cite{Yu_2023}. 
We show the mean values of scale heights and the standard deviation for each selection method and fit in Figure~\ref{fig:scaleheights}. 

Our scale height values are consistent with the thick disk scale height in the Milky Way \citep{Ma_2017}. This is expected since our simulations contain no gas component or star formation, which tend to populates the \textit{thin} disk \citep{Ma_2017}. 
The exponential-fit scale height drops by   from ~1.65 kpc for CDM to ~1.38 (1.5, and 1.59)  kpc for SIDM~5 (SIDM~2.5, and SIDM~1). 
Independently of the fit and selection of stars, the scale height drops by $4 \%$, $9 \%$, and $~17\%$ for SIDM~1, SIDM~2.5, and SIDM~5 relatively to CDM. 

\begin{figure}[]
    \centering
    \includegraphics[width=\columnwidth]{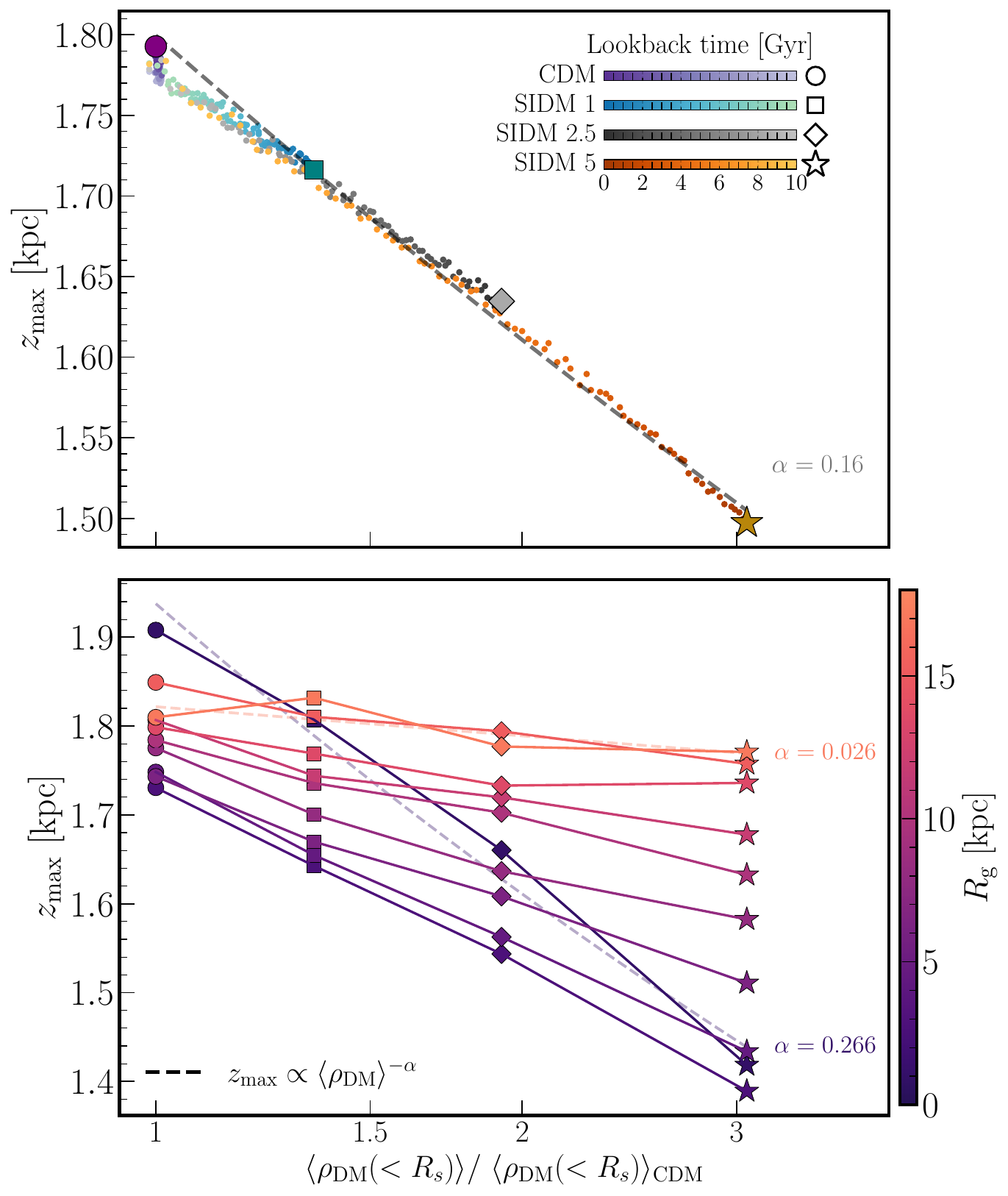}
    \caption {\textit{Top:} Median $z_{\rm{max}}$ for each model as a function of the average enclosed DM density within the disk scale radius, normalized by the corresponding CDM value. Points trace individual time steps, with lookback time indicated by the colorbars and large markers denoting the present day. The dashed line shows the best fit to the present-day values, with $\alpha = 0.160$; the adiabatic prediction, $\alpha = 0.25$, is shown for comparison. \textit{Bottom:} Present-day mean $z_{\rm max}$ separated into bins of guiding radius $R_{\rm g}$, indicated by the colorbar. The $z_{\rm max}$--density relation becomes progressively shallower with increasing $R_{\rm g}$, while the innermost bin, $R_{\rm g}\sim1.8$ kpc, yields $\alpha = 0.266$, consistent with the adiabatic prediction.
    \label{fig:zmax}}
\end{figure}

\begin{figure}[]
    \centering
    \includegraphics[width=\columnwidth]{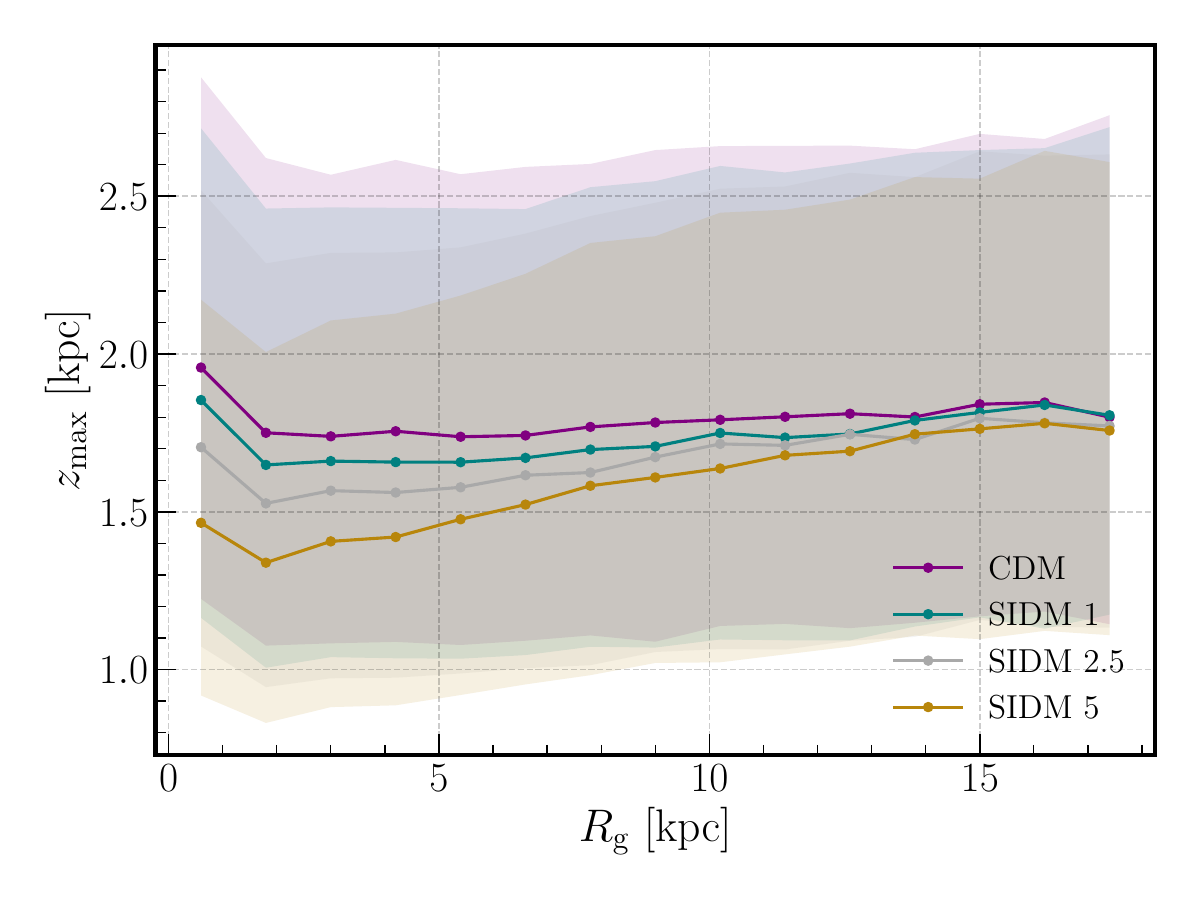}
    \caption{Median $z_{\rm max}$ of stars in the disk as a function of the guiding radius at present day. We show the $1\sigma$ band. CDM is shown in purple, SIDM~1 in teal, SIDM~2.5 in gray, and SIDM~5 in yellow. Disks in higher-cross-section models exhibit more prominent flaring because core collapse is concentrated in the inner halo, within $\sim2$ kpc. }
    \label{fig:flaring}
\end{figure}

While these results demonstrate the expected overall trend of decreasing scale height with increasing cross section, scale height is a fit to the entire disk and thus does not capture how the vertical thickness of different radial segments of the disk change under evolution of a radially dependent potential. A more revealing investigation would be to examine how the thickness evolves as a function of radius. In Figure~\ref{fig:zmax} and Figure~\ref{fig:flaring}, we explore how the maximum vertical extent of a stellar orbit, $z_{\rm max}$, evolves as a function of guiding radius $R_g$\footnote{We use \texttt{agama} \citep{vasiliev2019agama})   to calculate $R_g$ as the radius of circular orbit with angular momentum $L=J_{\phi}$ in the equatorial plane.}. This allows us to see how the simulations deviate from the toy model described by Eq.~\ref{eq:H_rhoc}.

We find it insightful to explore the evolution of $z_{\rm max}$ as a function of average DM density within the disk scale radius, $\langle \rho_{\rm DM} (R<R_s)\rangle$\footnote{Measuring the density in a narrow range of radii approximates a flat density in that range, thus adhering better to the toy model.}.
Using Eq.~\ref{eq:zmax_versus_rho} as a template, we fit the data to
\begin{equation}\label{eq:zmax}
    z_{\rm max} \propto \langle\rho_{\rm DM} (<R_s) \rangle^{-\alpha},
\end{equation}
for each $R_g$ bin (in the bottom panel of Fig.~\ref{fig:zmax}) and for all $R_g$ combined (in the top panel of Fig.~\ref{fig:zmax})). The fits are shown by dashed lines.

The top panel of Fig.~\ref{fig:zmax} shows the disk-averaged response of $z_{\rm max}$ to host halo density. At each time step, we plot the median $z_{\rm{max}}$ of each model against the average enclosed DM density within the disk scale radius, normalized by the corresponding CDM value. Lookback time is indicated by the colorbars, while the large markers denote the present-day values. Fitting Eq.~\ref{eq:zmax} to these present-day values, we find a best-fit slope of $\alpha = 0.160$, shallower than the toy model prediction ($\alpha = 0.25$), for a flat density profile undergoing adiabatic contraction  (Sec.~\ref{sec:adiabatic_compression}). This deviation is expected: the disk-averaged measurement includes stars outside the collapsing core, where the local DM density evolves far less strongly than the central density against which we fit, and the vertical dynamics are increasingly dominated by the disk's own self-gravity. Nevertheless, the monotonic decrease of the disk-averaged $z_{\rm max}$ with $\langle \rho_{\rm DM}(<R_s) \rangle$ demonstrates that the increased central density leaves an imprint on the thickness of the disk as a whole.

The bottom panel of Fig.~\ref{fig:zmax} separates the \textit{present-day} average $z_{\rm{max}}$ in bins of guiding radius, indicated by the colorbar, with marker shapes corresponding to the same models shown in the top panel. A clear radial gradient emerges: as guiding radius increases, the relation between $z_{\rm max}$ and the average DM density becomes progressively shallower. For the innermost orbit, with $R_{\rm g}$ $\sim 1.8$ kpc, we find $\alpha=0.266$, consistent with the adiabatic prediction. These stars orbit within $R_g$ comparable to the core radius (Figure~\ref{fig:tc}) where the density profile is approximately flat and the harmonic approximation of Eq.~\ref{eq:restoring_force} holds. $\alpha$ declines monotonically with $R_g$ and orbits with the outermost guiding radii (beyond $\sim 10$ kpc) are barely changed ($\alpha = 0.026$). This is a signature of a centrally confined process: SIDM core collapse redistributes mass only within the inner halo, while the density of the envelope remains effectively constant across DM models beyond $\sim 10$ kpc as we showed in Figure~\ref{fig:densitiesanddispersion}. Stars at large $R_g$ therefore feel a constant potential across the different cross sections even as the central density grows and their vertical structure remains close to constant. 

Core collapse, as the name implies, is a phenomenon associated with the inner radii of the DM halo. A clear consequence of radial-dependent potential variation, described in the previous section, is a radial-dependent disk response. In Figure ~\ref{fig:flaring}, we bin the disk orbits by guiding radius $R_g$ and then plot the median $z_{\rm max}$ in each bin. We demonstrate that our $N$-body simulations have a stronger disk thinning at lower radii than at higher radii, creating a `flaring' effect whereby the disk appears thicker at the edges. The compactification of the disk at the inner radii is much more prominent for higher cross sections; the SIDM 5 simulation showing a reduction by $\sim 25\%$ for the smallest guiding radii, compared with $5\%$ reduction for SIDM 1. 
The disk thins where the halo collapses (within $\sim 2$ kpc) and retains its initial thickness where it does not (beyond $\sim 10$ kpc), see Figure~\ref{fig:densitiesanddispersion}. This flared vertical profile grows with increasing $\sigma/m$ allowing the translation of the SIDM cross section and gravothermal evolution into a novel observable: the vertical structure of the stellar disk.

\section{Conclusions}\label{sec:conclusions}

We have presented $N$-body simulations of MW-size galaxies evolved in isolation for 10 Gyr with CDM and SIDM ($\sigma/m = 1.0, \,  2.5, \, 5.0 \,\rm{ cm}^2/\rm{g}$), each with and without a baryonic disk and bulge. Our simulations isolate the mutual effects between the baryonic component and the gravothermal evolution of the halo with SIDM. 

We find that the baryonic disk and bulge heat the core of the halo and invert its inner temperature gradient. While the temperature gradient in the DMO simulations drives heat inward and expands the core, the presence of the disk and bulge inverts the heat flow within a few Gyr (Figure~\ref{fig:densitiesanddispersion}). This speeds up the core expansion phase and accelerates the core collapse timescale $t_{\rm coll}$ by more than an order of magnitude: from analytic DMO core collapse timescales $t_0 \sim \mathcal{O}(10^3)$~Gyr, to our new baryon-inclusive core collapse timescale $t_{\rm coll} \sim \mathcal{O}(10)$~Gyr. 
The SIDM~5 halo reaches core collapse within the lifetime of the MW, SIDM~2.5 and SIDM~1 collapse within a few, while their DMO counterpart never leaves the expansion phase (Figure~\ref{fig:tc}).

We describe this relative acceleration by a virial-plus-adiabatic-contraction argument. We find that $t_{\rm coll}/t_0~\approx~(1+x)^{-3/2}$ (Eq.~\ref{eq:analytic}) where $x = M_{\rm b}/M_{\rm DM}$ showing that the relative acceleration is set by the baryon-to-DM mass $x$ ratio, while the absolute timescale retains its $(\sigma/m)^{-1}$
 scaling. We solve Poisson's equation for a flattened baryonic distribution and show that the disk geometry contributes a secondary order-unity enhancement to this acceleration. This enhancement is more pronounced in thinner disks and at high $x$ (Figure~\ref{fig:poisson}).

We then show that the contraction of the halo causes the disk to thin, by reducing the maximum vertical distance of the orbits as $z_{\rm max} \propto \rho^{-1/4}$. We recover this scaling in the simulation within 2 kpc, comparable to the core radius. The scaling holds in the inner halo because the density profile inside the core is approximately flat for much of the evolution, so the potential is well described by the harmonic form of Eq. ~\ref{eq:restoring_force}. The stars within the innermost guiding radii sample a single, slowly evolving midplane density, satisfying the adiabatic assumptions of Section~\ref{sec:adiabatic_compression}. For the full disk, however, the median values of $z_{\rm max}$ deviate from the theoretical expectations, and are found to be of $z_{\rm max} \propto \langle\rho(<R_s)\rangle^{-0.16}$, showing that the outermost orbits are essentially unaffected (Figure~\ref{fig:zmax}).  
This deviation is expected to be due to a number of reasons including the disk's own self-gravity, the initially core expansion that our model does not capture, and the non flat core density during collapse. 

At the final snapshot, the disk scale height deceases by $\sim 4\%, 9\%, 17\%$ for SIDM~1, SIDM~2.5 and SIDM~5 relative to CDM, independently of the star selection and the assumed vertical profile (Figure~\ref{fig:scaleheights}). This is explained by the increased DM potential towards the center of the halo. 

Since core collapse is centrally confined, we find that the thinning is both cross section and radius-dependent, which causes disk flaring for higher $\sigma/m$ values (Figure~\ref{fig:flaring}). Therefore, the disk thins from inside out and its vertical structure can encode SIDM cross section information. Because disk heating and stability influence star formation and chemical enrichment, these effects may leave observable structural and chemical imprints in the disk. 

The results presented in this work are obtained in a controlled setting. Our halos evolve in isolation, with a fixed baryonic mass and baryon-to-DM ratio, with velocity-independent cross sections, and without gas, star formation, feedback, or mergers. Consequences of these simplification include: 

First, our total baryonic mass $M_{\rm{disk}} + M_{\rm{bulge}} = 4.1 \times 10^{10} {\rm M}_\odot$, remains below the present day value for the Milky Way's baryonic mass $\sim 6 \times 10^{10} {\rm M}_\odot$ \citep{Bland_Hawthorn_2016}. We argue that this makes our collapse timescales conservative since a larger baryonic mass raises $x$ in Eq.~\ref{eq:analytic} and therefore could shorten the core-collapse time further \citep[e.g.][]{Zhong_2023}.

Second, our disk is more representative of the MW's thick-disk \citep{Ma_2017}. Therefore, our absolute scale heights are not directly comparable to the MW disk, but within our setup we see a decrease in the scale height as a function of $\sigma/m$. Including gas and stellar formation would allow for a thin disk to form, which would be interesting to investigate in the light of the SIDM thinning and flaring of the disk. 

Third, our isolated setup omits processes that act on the collapse timescales in opposite directions. Mergers suppress core collapse \citep{silverman_2026}, while the continued cooling and contraction of baryons in a cosmological setting accelerates it by deepening the central potential \citep{Gnedin_2004}. Our isolated timescales capture neither effects. Forthcoming follow-up work to this paper will study these timescales in the presence of a merger like the Gaia Sausage Enceladus \citep{Helmi_2018, Belokurov_2018}. 

Finally, to generalize our single halo mass and baryon-to-DM ratio results and turn them into an observable of SIDM, future work will investigate population statistics; we will consider different initial conditions and sample a set of varied halo masses and baryon-to-DM enclosed mass ratios. 

In conclusion, baryons fundamentally reshape the collapse landscape: MW-mass halos begin core-collapse within a Hubble time for cross sections as low as $\sigma/m = 1\,{\rm cm^2/g}$, driven not just by baryonic mass but by disk geometry; thinner disks collapse faster at fixed mass. The vertical structure of the stellar disk has long been a fossil record of a galaxy's merger history, star formation, and quenching. Here, we add another layer to that record: as the halo undergoes gravothermal collapse, the disk thins and flares, imprinting the DM self-interaction cross section on the stellar distribution of galaxies, providing a new probe of the particle nature of dark matter.

\software{This work made use of 
\texttt{GalIC}~\citep{Yurin_2014},\texttt{Gizmo}~\citep{Gizmo} for the $N$-body simulations,
\texttt{matplotlib}~\citep{mplplt_2007}, \texttt{NumPy}~\citep{numpy_2011} \texttt{pandas}~\citep{pandas_2010}, \texttt{SciPy}~\citep{Jones_2001}, \texttt{Agama}~\citep{vasiliev2019agama} for the analysis, and \texttt{OverCite}~\citep{Shariat2026}, an in-editor citation tool for \LaTeX.}

\facilities{The authors acknowledge the MIT Office of Research Computing and Data (\url{https://orcd.mit.edu/}) for providing high performance computing resources that have contributed to the research results reported within this paper.}

\begin{acknowledgments}
We thank Maya Silverman and Nathaniel Starkman for insightful discussions about core collapse timescales, and Hanyuan Zhang and the Cambridge Streams group for useful discussions regarding galactic disks. We also thank Tri Nguyen and Nora Shipp for their helpful feedback at the early stages of this project. 
The work of Z.M. is supported by the MIT Dean of Science Graduate Fellowship.
A.D. is supported  by the European Union (ERC, BeyondSTREAMS, 101115754) grant. Views and opinions expressed are however those of the authors only and do not necessarily reflect those of the European Union or the European Research Council. Neither the European Union nor the granting authority can be held responsible for them. 
The work of G.H. is supported by the Neutrino Theory Network Fellowship with contract number 726844.
A.H. is supported by NSF award 2307788. A.H acknowledges that this manuscript has been authored by Fermi Forward Discovery Group, LLC under Contract No. 89243024CSC000002 with the U.S. Department of Energy, Office of Science, Office of High Energy Physics. A.H. acknowledges that this material is based upon work supported by the U.S. Department of Energy, Office of Science, Office of Workforce Development for Teachers and Scientists, Office of Science Graduate Student Research (SCGSR) program. The SCGSR program is administered by the Oak Ridge Institute for Science and Education (ORISE) for the DOE. ORISE is managed by ORAU under contract number DESC0014664. All opinions expressed in this paper are the author’s and do not necessarily reflect the policies and views of DOE, ORAU, or ORISE. A.H. would like to thank the Kavli Institute for Cosmological Physics~(KICP) at the University of Chicago for their hospitality, where part of this work was conducted. 
L.N. is supported by the Sloan Fellowship, 
the NSF CAREER award 2337864, and NSF award 2307788. 
L.N. also gratefully acknowledges the continued support of the Adam J. Burgasser Endowed Chair of Astrophysics at MIT, which sustained many long hours of writing and revision of this manuscript.

\end{acknowledgments}

\begin{contribution}

Z.M. ran the simulations, performed the analysis, and drafted the manuscript. E.Y.D. derived the toy model for disk thinning and wrote Section~\ref{sec:adiabatic_compression}. A.D. co-wrote the analysis code and contributed to the development and validation of the simulations and analysis pipeline. G.H. derived the Poisson solution, implemented the corresponding solver, and wrote section~\ref{sec:densityincrease}. A.H. implemented time-stepping revisions to the simulation code. L.N. conceived and supervised the project, and acquired funding and computational resources. All authors contributed to writing, reviewing, and editing the manuscript.

\end{contribution}

\bibliography{intro}{}
\bibliographystyle{aasjournalv7}

\clearpage
\appendix

\setcounter{equation}{0}
\setcounter{figure}{0} 
\setcounter{table}{0}
\renewcommand{\theequation}{A\arabic{equation}}
\renewcommand{\thefigure}{A\arabic{figure}}
\renewcommand{\thetable}{A\arabic{table}}
\renewcommand*{\theHfigure}{\thefigure}
\renewcommand*{\theHtable}{\thetable}
\renewcommand*{\theHequation}{\theequation}

\section{Convergence testing}
\label{app:convergence}

Resolving gravothermal core collapse in $N$-body simulations is nontrivial \citep{Yang_2022}. A timestep that is adequate for the most of the halo can still fail to resolve the self-interactions in high density regions. In this appendix we present the convergence tests that set the time-stepping parameters used throughout this work.

To resolve SIDM scatterings in high density regions in simulations, the time step size in \texttt{GIZMO} is set by two parameters (see \cite{Palubski_2024} for details): 
\begin{itemize}
    \item \textbf{The tolerance parameter $\eta$:} a dimensionless number describing the fraction of the force-softening length that the particle is allowed to move in a given time step.
    \item \textbf{The maximum probability of scattering $\kappa$:} in a given time-step each pair of particles is allowed to interact at a probability of $P_{ij} < \kappa$. 
\end{itemize}
While the values of $\kappa$ and $\eta$ that ensure convergence of the core density $\rho_{\rm c}(t)$ have been established for dwarf galaxies \citep{Palubski_2024,mace2024convergencetestsselfinteractingdark,engelhardt2026marvelouslydarkdensityprofile,silverman_2026}, we perform our own convergence tests to determine the appropriate time-stepping parameters for our MW-size galaxy.

\begin{figure}[th]
    \centering
    \includegraphics[width=0.48\textwidth]{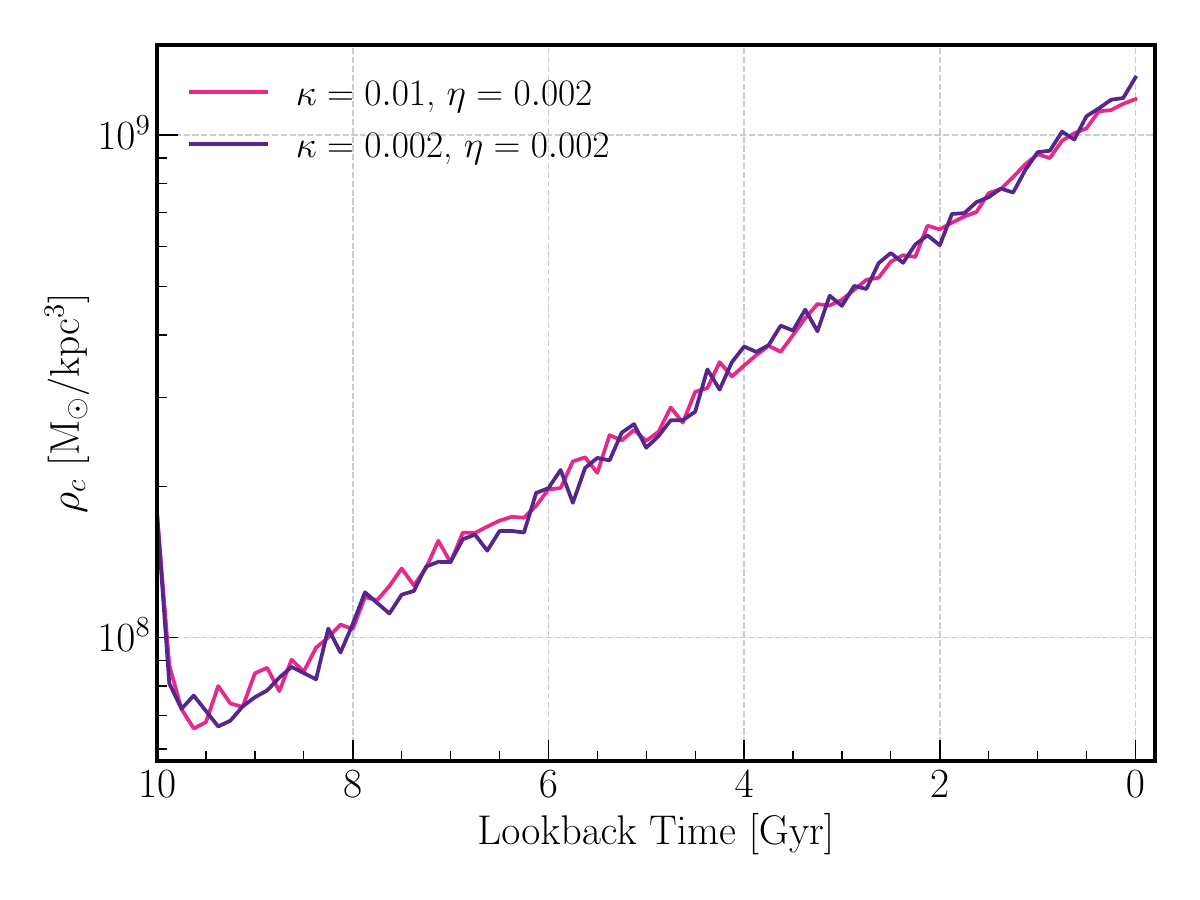}
    \includegraphics[width=0.48\textwidth]{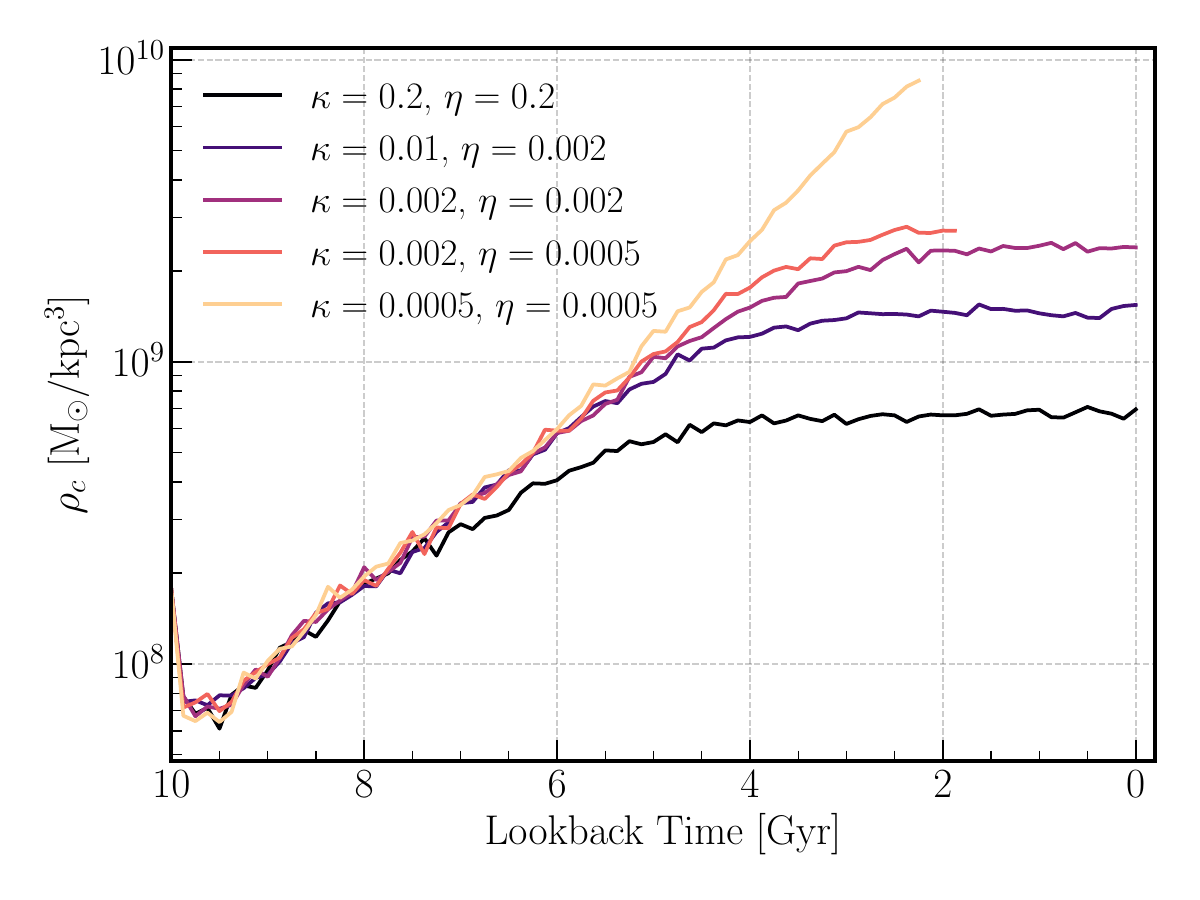}
    \caption{\textit{Left:} Core density vs time for our MW simulations with baryons and cross section $\sigma/m=5~\mathrm{ cm}^2/\mathrm{g}$ we vary $\eta$ and $\kappa$ and find that $\rho_{\rm c}$ converges.
    \textit{Right:} Core density vs time for our MW simulations with baryons and cross section $\sigma/m=10~\mathrm{ cm}^2/\mathrm{g}$ we vary $\eta$ and $\kappa$ but the simulations appear to stall after a few Gyrs.
    }
    \label{fig:SIDM5}
\end{figure}

The left panel of Figure~\ref{fig:SIDM5} shows $\rho_{\rm c}(t)$ for the $\sigma/m = 5~\mathrm{cm}^2/\mathrm{g}$, run across several choices of the time stepping parameters. We see that the density has converged for $\kappa = 0.01$ and $\eta=0.002$, i.e. lower values of $\kappa,\eta$ (and subsequently finer timesteps) do not alter the core density evolution. 

We further plot the parameter sweep for $\sigma/m = 10~\rm{cm}^2/\rm{g}$ run in the right panel of Figure~\ref{fig:SIDM5}. We see that the core density flattens, as observed in \cite{Palubski_2024} for values of $\kappa =0.02$ for halos of mass $M_{\rm{DM}}~=~9.90~\times ~10^{11}~{\rm M}_\odot $, and thus we are unable to confirm if the density converges for our lowest $\kappa,\eta$ runs due to computational cost. We speculate that a similar numerical stalling may contribute to the non-collapsing isothermal cores reported at $\sigma/m =10 ~\mathrm{ cm}^2/\mathrm{g}$ in the FIRE suite in \cite{Sameie_2021}, though differences in setup (cosmological environment, hydrodynamics) preclude a direct comparison. Novel implementations of the SIDM scattering implementations are being developed \cite[see e.g.,][]{Zier:2026esu}, which could explore cross sections that the current time stepping schemes cannot resolve.

\clearpage

\section{Analytic Core Collapse Times}
\label{app:corecollapse}

\setcounter{equation}{0}
\setcounter{figure}{0} 
\setcounter{table}{0}
\renewcommand{\theequation}{B\arabic{equation}}
\renewcommand{\thefigure}{B\arabic{figure}}
\renewcommand{\thetable}{B\arabic{table}}
\renewcommand*{\theHfigure}{B\thefigure}
\renewcommand*{\theHtable}{B\thetable}
\renewcommand*{\theHequation}{B\theequation}

In section~\ref{sec:corecollapsetimesinsim}, we compute the DMO core-collapse time $t_0$ using the local relaxation time (Eq.~\ref{eq:t_relax}) evaluated on the simulation density and velocity dispersion profiles at $R = 10$ kpc. This choice ties the
timescale to the radius at which the baryons measurably alter the halo density profile (Section~\ref{sec:coreexp}).
Other works instead express these timescales through the parameters of an NFW fit to the initial halo \citep{Essig_2019, silverman_2026} using
\begin{equation}\label{eq:timescale}
    t_{\rm c} \approx \frac{150}{C}\, \frac{1}{r_s\, \rho_s\, \sigma/m}\, \frac{1}{\sqrt{4\pi G \rho_s}} ,
\end{equation}
with $r_s$ and $\rho_s$ the scale radius and density of the halo when fitted to an NFW profile, and $C= 0.60–0.84$ the heat transfer constant \citep{Essig_2019,Nishikawa_2020, Palubski_2024, Yang_2024}. Both Eq.~\ref{eq:t_relax} and Eq.~\ref{eq:timescale} encode
the same physics, $t_0 \propto(\rho\,\sigma_v\,\sigma/m)^{-1}$, but differ in where the halo is sampled: Eq.~\ref{eq:t_relax} at a radius we choose (10 kpc) and  Eq.~\ref{eq:timescale} at the characteristic radius fixed by the NFW fit.

Similarly to \cite{silverman_2026}, we adopt the geometric mean of this range, C = 0.71.
The halos reach the end of core expansion at $t_{\rm core,0} = 50 \, t_{c,0}$ and the end of core collapse at $t_0=400 \, t_{c,0}$ \citep{Outmezguine:2022bhq, Palubski_2024, silverman_2026}.

\begin{table*}[h]
    \centering
    \renewcommand{\arraystretch}{1.5}
    \begin{tabular}{lcccc}
        \hline
        Simulation & $t_{\rm core,0}$ & $t_{\rm core}$ & $t_0$ & $t_{\rm coll}$ \\
        \hline
        SIDM 1   & $172.34 \pm 1.0$ & $3.38 \pm 2.3$ & $1378.76 \pm 8.0$ & $49.9^{+39.6}_{-12.7}$ \\
        SIDM 2.5 & $68.94 \pm 0.4$  & $1.12 \pm 0.4$ & $551.50 \pm 3.2$  & $17.4^{+1.7}_{-0.0}$ \\
        SIDM 5   & $34.47 \pm 0.2$  & $0.88 \pm 0.1$ & $275.75 \pm 1.6$  & $8.6^{+0.1}_{-0.0}$ \\
        \hline
    \end{tabular}
    \caption{We calculate the theoretical predictions for DMO runs for end of core expansion timescale $t_{\rm core,0}$ and core collapse $t_0$ using NFW fits and Eq.~\ref{eq:timescale}. We identify the end of core expansion $t_{\rm core}$ and infer the end of core collapse $t_{coll}$ for the Halo+Baryons simulations by smoothing and fitting the core densities $\rho_{\rm c}$. The times reported are in Gyr.}\label{tab:nfw_timescales}
\end{table*}

Table~\ref{tab:nfw_timescales} lists the predicted DMO timescales $t_{\rm core,0}$ and $t_0$ using Eq.~\ref{eq:timescale} alongside the measured $t_{\rm core}$ and inferred $t_{\rm coll}$ of the Halo+Baryons runs, which are unchanged from Table~\ref{tab:times}. The NFW-based $t_{\rm core,0}$ and $t_0$ are shorter than the values we show in Table~\ref{tab:times} by a common factor of $\sim 0.6$ and $0.7$ for all cross sections. This systematic decrease is due Eq.~\ref{eq:timescale} evaluating the scattering rate with the NFW scale quantities rather than locally. We find that the fitted product $\rho_s v_s$ that sets the scattering rate is $\sim2$ times its local value. The two agree if Eq.~\ref{eq:corecollapse} is instead evaluated at $r \sim 7$~kpc. The quoted uncertainties on $t_{\rm core,0}$ and $t_0$ come from the uncertainty in the fitted NFW parameters (because the halos are initialized as Hernquist profiles), which we propagate in quadrature with the uncertainty on $t_{\rm coll}$ when forming the ratio $t_{\rm coll}/t_0$ below.

\begin{figure}[th]
    \centering
    \includegraphics[width=0.5\columnwidth]{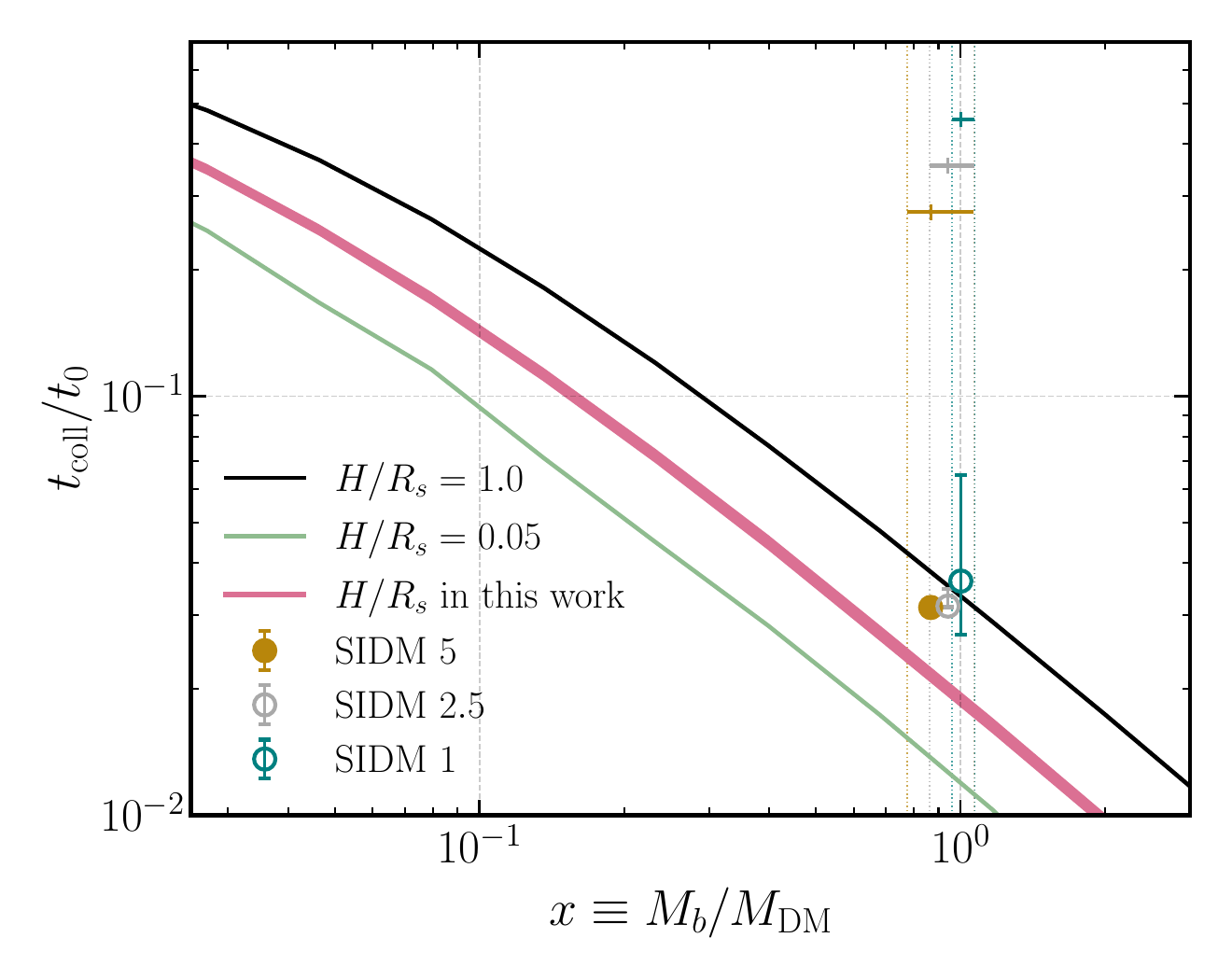}
    \caption{Poisson equation solution (Eq.~\ref{eq:poisson}) for disk geometry. We show a solution for a thick disk ($H/\ R_s=1$) in black, a thin disk ($H/\ R_s=0.05$) in green, and the range for the disks in our simulations ($H/\ R_s=[0.380, 0.442]$) in pink. $t_{\rm{coll}}$ calculated from the simulations is marked by yellow (SIDM 5), gray (SIDM 2.5) and teal (SIDM 1) circles, where the filled circles mark a time that was measured using the simulation data and a hollow circle marks a time that was obtained through fits} 
    \label{fig:eq14figure}
\end{figure}

Figure~\ref{fig:eq14figure} reproduces Figure~\ref{fig:poisson} with $t_0$ calculated using Eq.~\ref{eq:timescale}. The curves show the Poisson-equation prediction for $t_{\rm coll}/t_0$ as a function of the baryon-to-DM mass ratio $x = M_{\rm b}/M_{\rm DM}$ for a thick disk ($H/R_s = 1$) in black, a thin disk ($H/R_s = 0.05$) in green, and the range spanned by our simulated disks ($H/R_s = [0.380, 0.442]$) in pink. The circles mark the $t_{coll}/t_0$ ratio for SIDM~1 (teal) SIDM~2.5 (gray) and SIDM~5 (yellow). Because $t_0$ is now shorter (while $t_{coll}$ is unchanged), each point is moved up relative to Figure~\ref{fig:poisson}, and lands closer to the thick-disk curve rather than on the pink band that corresponds to our actual disk geometry. While the extended tail of the bulge partially compensates for the missing gas mass, at $\sim 7$~kpc, it does not effectively capture the flattened disk geometry discussed in Section~\ref{sec:densityincrease}. Therefore, we find that the core collapse timescales at this radius are more consistent with $H/R_s = 1.0$. It is important to note that the ordering of the points and the order-of-magnitude acceleration relative to the DMO prediction are still the same under either definition.

\end{document}